\documentclass[prx,aps,showpacs,twocolumn,superscriptaddress,longbibliography]{revtex4-2}
\usepackage{graphicx}
\usepackage{mathrsfs}
\usepackage{amssymb}
\usepackage{amsmath}
\usepackage{physics}
\usepackage{bbding}
\usepackage{xcolor}
\usepackage[colorlinks=true,linkcolor=blue,anchorcolor=red,citecolor=blue,urlcolor=blue]{hyperref}

\begin{document}

\title{Linear Magnetoresistance of 2D Massless Dirac Fermions in the Quantum Limit}

\author{Xiao-Bin Qiang}
\email{These authors contributed equally to this work.}
\affiliation{State Key Laboratory of Quantum Functional Materials, Department of Physics, and Guangdong Basic Research Center of Excellence for Quantum Science, Southern University of Science and Technology (SUSTech), Shenzhen 518055, China}
\affiliation{Division of Physics and Applied Physics, School of Physical and Mathematical Sciences, Nanyang Technological University, 21 Nanyang Link, 637371, Singapore}

\author{Han-Yi Xu}
\email{These authors contributed equally to this work.}
\affiliation{State Key Laboratory of Quantum Functional Materials, Department of Physics, and Guangdong Basic Research Center of Excellence for Quantum Science, Southern University of Science and Technology (SUSTech), Shenzhen 518055, China}

\author{Ren-Jie Tong}
\email{These authors contributed equally to this work.}
\affiliation{State Key Laboratory of Quantum Functional Materials, Department of Physics, and Guangdong Basic Research Center of Excellence for Quantum Science, Southern University of Science and Technology (SUSTech), Shenzhen 518055, China}

\author{Shuai Li}
\affiliation{Hubei Engineering Research Center of Weak Magnetic-field Detection, Department of Physics, China Three Gorges University, Yichang 443002, China}

\author{Zi-Xuan Gao}
\affiliation{State Key Laboratory of Quantum Functional Materials, Department of Physics, and Guangdong Basic Research Center of Excellence for Quantum Science, Southern University of Science and Technology (SUSTech), Shenzhen 518055, China}

\author{Peng-Lu Zhao}
\email{Corresponding author: zhaoplu@gmail.com}
\affiliation{Quantum Science Center of Guangdong-Hong Kong-Macao Greater Bay Area (Guangdong), Shenzhen 518045, China}

\author{Hai-Zhou Lu}
\email{Corresponding author: luhz@sustech.edu.cn}
\affiliation{State Key Laboratory of Quantum Functional Materials, Department of Physics, and Guangdong Basic Research Center of Excellence for Quantum Science, Southern University of Science and Technology (SUSTech), Shenzhen 518055, China}
\affiliation{Quantum Science Center of Guangdong-Hong Kong-Macao Greater Bay Area (Guangdong), Shenzhen 518045, China}

\date{\today}

\begin{abstract}
Linear magnetoresistance is a hallmark of 3D Weyl metals in the quantum limit. Recently, a pronounced linear magnetoresistance has also been observed in 2D graphene [\href{https://doi.org/10.1038/s41586-023-05807-0}{Xin \emph{et al}., Nature \textbf{616}, 270 (2023)}]. However, a comprehensive theoretical understanding remains elusive. By employing the self-consistent Born approximation, we derive the analytical expressions for the magnetoresistivity of 2D massless Dirac fermions in the quantum limit. Notably, our result recovers the minimum conductivity in the clean limit and reveals a linear dependence of resistivity on the magnetic field for Gaussian impurity potentials, in quantitative agreement with experiments. These findings shed light on the magnetoresistance behavior of 2D Dirac fermions under ultra-high magnetic fields.
\end{abstract}
\maketitle

%\textit{\textcolor{blue}{Introduction.}--}
\section{Introduction}
Magnetoresistance refers to the variation in resistance of materials in response to an applied magnetic field. In most conventional materials, the magnetoresistance typically follows a positive quadratic dependence on the magnetic field~\cite{Ashcroft76book}. Deviations from this standard behaviour can reveal exotic properties of materials, and provides a powerful tool for exploring underlying phenomena.

In a weak magnetic field, negative (positive) magnetoresistance is a hallmark of weak localization (antilocalization)~\cite{Bergmann84pr,PatrickLee85rmp,Ando02prl,Lu11prl,Lu13prl,Lu14prl,FuB19prl,WangHW20prl}. In a moderate magnetic field, negative magnetoresistance can arise from the chiral anomaly~\cite{Spivak13prb,Burkov14prl,FangZ15prx,Ong15science} or Berry curvature~\cite{WangJ12nanores,Wiedmann16prb,Yoichi17nc,Lu17prl,ChenJH23sb}. In a higher magnetic field, Shubnikov–de Haas oscillation of magnetoresistance serves as a critical evidence for the quantum Hall effect~\cite{Klitzing80prl,Geim05nature,XiuFX19nature,ZhangLY19nature}. Besides, in an ultra-high magnetic field, where only the zeroth Landau level is occupied (quantum limit), linear magnetoresistance emerges as a hallmark of 3D Weyl metals~\cite{Abrikosov98prb,HuJ08nm,HeLP14prl,Coldea15prl,Ong15nm,YanBH15np}. Notably, recent experiment~\cite{Geim23nature} has reported a pronounced linear magnetoresistance in graphene under the quantum limit condition. Due to 2D electronic structure of graphene, where the energy spectrum is quantized into a series of Landau levels in a strong magnetic field, existing theoretical framework based on the first-order Born approximation~\cite{Abrikosov98prb,LiS23prb} fails to model this 2D system. This highlights a theoretical gap that warrants further investigation.

\begin{figure}[htbp]
\centering 
\includegraphics[width=0.5\textwidth]{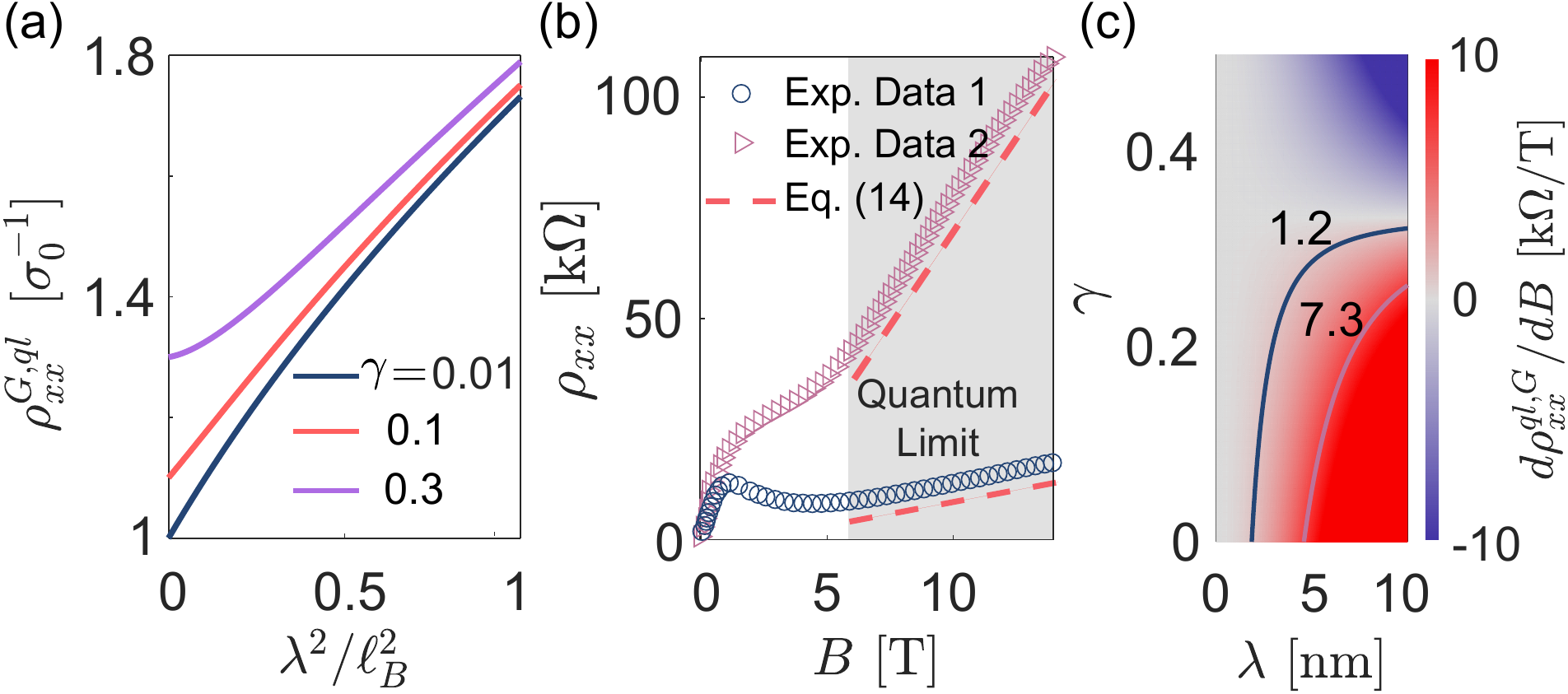}
\caption{(a) Calculated magnetoresistivity [Eq.~\eqref{Eq: rho_G}] as a function of the dimensionless parameter $\lambda^2/\ell_B^2$  for various impurity strengths $\gamma$, where $\lambda$ is the decay length of the Gaussian potential, $\ell_B = \sqrt{\hbar/(eB)}$ is the magnetic length, and $\sigma_0=e^2/(2\sqrt{2}\pi h)$ is the minimal conductivity of 2D massless Dirac fermions~\cite{Ludwig94prb,Ziegler97prb,Mudry07prb,Cserti07prb,Ziegler07prb,Dimi22prl}. (b) Comparison between linear magnetoresistivity in the quantum limit [Eq.~\eqref{Eq: rho_G2}] and experimental data~\cite{Geim23nature}. (c) Phase diagram for the slope of the magnetoresistivity $d\rho_{xx}^{G,ql}/dB$ as a function of the decay length $\lambda$ and impurity strength $\gamma$. Here, $\gamma$ cannot be zero unless $\lambda$ approaches zero first. Two inset solid curves indicate the ranges of parameters consistent with the experimental slopes (1.2 and 7.3  k$\Omega$/T) shown in panel (b).}
\label{Fig: exp}
\end{figure}

In this work, we present a comprehensive study of the magnetoresistance of 2D massless Dirac fermions in the quantum limit. Starting from linear-response theory~\cite{Mahan90book} and treating impurity scattering within the self-consistent Born approximation~\cite{Ando98jpsj,Ando02prb,Mirlin06prb}, we derive analytical expressions for the longitudinal resistivity $\rho_{xx}(B)$. Our analysis includes three distinct types of impurity potentials including $\delta$-function potential, Gaussian potential, and Yukawa potential, each resulting in a different dependence of resistivity on the magnetic field. When the impurity strength approaches to zero, the derived results consistently converge to the well-established minimum conductivity of 2D massless Dirac fermions~\cite{Ludwig94prb,Ziegler97prb,Mudry07prb,Cserti07prb,Ziegler07prb,Dimi22prl}. Notably, we find that the Gaussian impurity potential can exhibit a linear magnetoresistivity whenever its decay length is smaller than the magnetic length. As shown in Fig.~\ref{Fig: exp}, our analytical results [Eqs.~\eqref{Eq: rho_G} and \eqref{Eq: rho_G2}] are quantitatively in agreement with experimental observations \cite{Geim23nature}.

%\textit{\textcolor{blue}{Model and Landau levels.}--} 
\section{Model and Landau levels}
\begin{figure}[htbp]
\centering 
\includegraphics[width=0.48\textwidth]{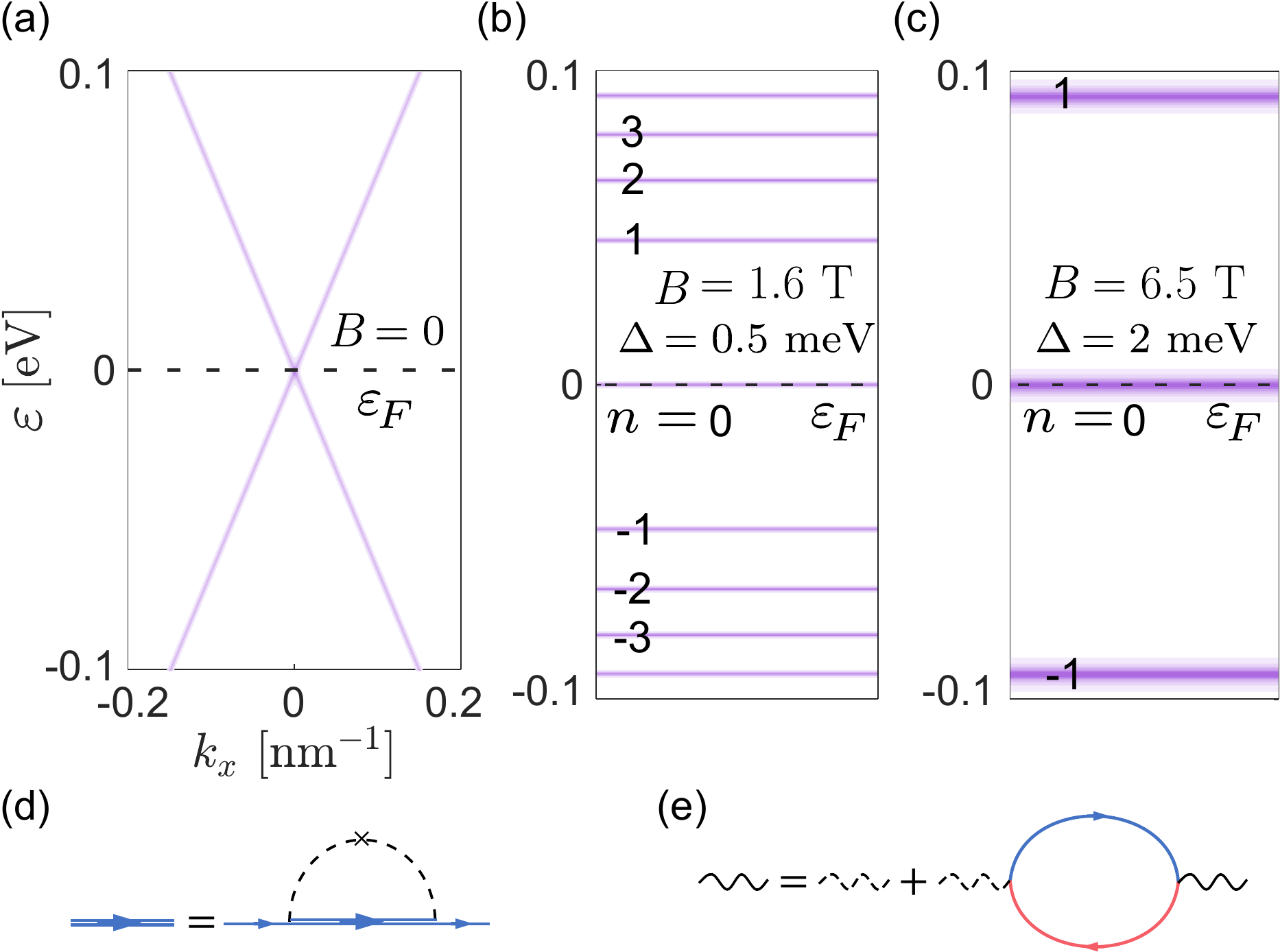}
\caption{Energy spectrum of massless Dirac fermions [Eq.~\eqref{Eq: model}]. (a) In absence of magnetic field, (b) Landau levels with magnetic field $B=1.6$~T (assuming a constant line-width $\Delta=0.5$~meV), and (c) Landau levels with $B=6.5$~T and $\Delta=2$~meV. The Landau levels are denoted by indices $n = 0,\pm1,\pm2,\cdots$. The dashed line represents the Fermi level $\varepsilon_F$. The model parameter is $v=0.65$~eV$\cdot$nm adopted from Ref.~\cite{Geim05nature}. Feynman diagrams for (d) the self-consistent Born approximation, and (e) the screened length $\kappa$ of the Yukawa potential. In these diagrams, the double solid line represents the dressed Green function, while the single solid line represents the bare Green function, and the dashed line indicates impurity scattering. The solid wavy line represents interaction within the random phase approximation, and the dashed wavy line represents to the bare Coulomb interaction.}
\label{Fig: spectra}
\end{figure}

The Hamiltonian of 2D massless Dirac fermions is ~\cite{Geim05nature,Shen17book}
\begin{equation}\label{Eq: model}
\mathcal{H}_0=v\mathbf{k}\cdot\boldsymbol{\sigma},
\end{equation}
where $v$ is the model parameter characterizing the Fermi velocity, $\mathbf{k}=(k_x,k_y)$ is the wave vector, and $\boldsymbol{\sigma}=(\sigma_x,\sigma_y)$ is the vector of Pauli matrices. This Hamiltonian serves as an effective model for a range of novel quantum materials, such as the surface states of topological insulators~\cite{Kane10rmp,ZhangSC11rmp} and graphene~\cite{Neto09rmp,DasSarma11rmp}. In the case of graphene, it represents the effective low-energy Hamiltonian near one of the two inequivalent valleys. The counterpart valley possesses an identical Landau-level spectrum and yields equivalent physical responses in the present context. Therefore, considering a single valley is sufficient for describing the magnetotransport phenomena discussed in this work. The energy spectrum of this model [Eq.~\eqref{Eq: model}] is given by $\varepsilon_{\pm} = \pm v k$, where $k=|\mathbf{k}|$. As shown in Fig.~\ref{Fig: spectra}(a), this linear energy spectrum is a defining characteristic of 2D massless Dirac fermions and underlies a variety of novel physical properties~\cite{DasSarma11rmp,Lu23prb,QinF25prb1,ChenR25jpcm,QinF25prb2}.

In a perpendicular magnetic field $\mathbf{B}=B\hat{z}$, the wave vector in the Landau gauge is modified to $\mathbf{k}=(k_x+eA_x/\hbar,i\partial_y)$ with vector potential $\mathbf{A}=-yB\hat{x}$, where $-e$ is the electron charge, and $\hbar=h/(2\pi)$ is the reduced Planck constant. Then, by introducing the ladder operators~\cite{Lu15prb,WangHW18prb} $a=-[(y-y_0)/\ell_B+\ell_B\partial_y]/\sqrt{2}$ and $a^\dagger=-[(y-y_0)/\ell_B-\ell_B\partial_y]/\sqrt{2}$ with guiding center $y_0=\ell_B^2k_x$ and magnetic length $\ell_B=\sqrt{\hbar/(eB)}$, the Hamiltonian [Eq.~\eqref{Eq: model}] becomes
\begin{equation}
\mathcal{H}_0=\begin{bmatrix}
0 &  (\sqrt{2}v/\ell_B) a\\
(\sqrt{2}v/\ell_B) a^\dagger & 0
\end{bmatrix}.
\end{equation}
With trial wave functions of the form $[0, |0\rangle]^\text{T}$ for $n=0$, and $[c_1|n-1\rangle, c_2|n\rangle]^\text{T}$ for $n\neq 0$, the eigenenergies are solved as (details can be found in Sec. SI of Supplemental Material~\cite{Supp})
\begin{equation}\label{Eq: LL}
\varepsilon_{n}=\pm\frac{1}{\ell_B}\sqrt{2v^2n},\quad n=0,1,2,\cdots.
\end{equation}
As shown in Figs.~\ref{Fig: spectra}(b) and (c), Eq.~\eqref{Eq: LL} represents a series of Landau levels for 2D massless Dirac fermions. Additionally, the wave functions are given by
\begin{equation}\label{Eq: WF}
\begin{aligned}
\psi_{k_x,\pm n}(\mathbf{r})&=\frac{e^{ik_x x}}{\sqrt{2L_x}}  
\begin{bmatrix}
\phi_{k_x,n-1}(y) \\
\pm\phi_{k_x,n}(y) \\
\end{bmatrix},\quad n=1,2,\cdots, \\
\psi_{k_x,0}(\mathbf{r})&=\frac{e^{ik_x x}}{\sqrt{L_x}}  
\begin{bmatrix}
0 \\
\phi_{k_x,0}(y)\\
\end{bmatrix},\quad n=0,\\
\end{aligned}
\end{equation}
where $L_x$ is the system length, $\phi_{k_x,n}(y)=e^{-\zeta^2/2}H_n(\zeta)/\sqrt{n!2^n\sqrt{\pi}\ell_B}$ with $\zeta=(y-y_0)/\ell_B$, and $H_n(\zeta)$ is the Hermite polynomial.

In the quantum limit, where only the zeroth Landau level ($\varepsilon_0=0$) is occupied, which implies that the Fermi energy is $\varepsilon_F=0$. This indicates that the quantum limit coincides with the charge neutrality point. This behaviour is fundamentally different from that of a conventional 2D electron gas~\cite{Landau30zfp}, where the lowest Landau level lies at a finite energy $\hbar\omega/2$, with $\omega$ denoting the cyclotron frequency.

%\textit{\textcolor{blue}{Magnetoresistivity in the quantum limit.}--}
\section{Magnetoresistivity in the quantum limit}
Following the standard linear-response theory~\cite{Mahan90book}, the longitudinal conductivity in the quantum limit is given by (see Sec. SII of~\cite{Supp})
\begin{equation}\label{Eq: sigma_AA}
\sigma_{xx}^{ql}=\frac{e^2 v^2}{8\pi\hbar\mathcal{V}}\sum_{k_x,n=\pm 1}\mathcal{A}_{k_x, 0} (\varepsilon_F)\mathcal{A}_{k_x, n}
(\varepsilon_F),
\end{equation}
where the superscript of $\sigma_{xx}^{ql}$ denotes the quantum limit, and
$\mathcal{V}$ is the system volume. The spectral function is given by $\mathcal{A}_{k_x,n} = 2\Delta_{k_x,n} /[(\varepsilon_F - \varepsilon_n)^2 + \Delta_{k_x,n}^2]$, where $\Delta_{k_x,n}$ is the line-width function. Figs.~\ref{Fig: spectra}(b) and (c) illustrate the impurity-dressed Landau levels [Eq.~\eqref{Eq: LL}] with help of the spectral function. The effect of impurity scattering is to broaden the Landau levels into finite widths.

The line-width function is generally written as $\Delta_{k_x, n}=-\text{Im}[\Sigma_{k_x, n}^r]$, where $\Sigma_{k_x, n}^r$ is the impurity scattering induced selfenergy. Under the self‑consistent Born approximation~\cite{Ando98jpsj,Ando02prb,Mirlin06prb}, $\Sigma_{k_x, n}^r$ can be calculated from the Feynman diagram in Fig.~\ref{Fig: spectra}(d) and is found as
\begin{equation}
\Sigma_{k_x,n}^r = \sum_{k_x', n'}\frac{|\langle k_x', n' |V(\mathbf{r})|k_x, n \rangle|^2}{\varepsilon_F - \varepsilon_{n'} - \Sigma_{n'}^r}.
\end{equation}
Here, the scattering matrix element is given by
\begin{equation}
\langle k_x', n'| V(\mathbf{r})|k_x, n \rangle = \int d \mathbf{r}
\psi_{k_x',n'}^{\dagger} (\mathbf{r}) V (\mathbf{r}) \psi_{k_x,n}
(\mathbf{r}),
\end{equation}
where $\psi_{k_x,n}(\mathbf{r})$ is the wave function of Landau level given by Eq.~\eqref{Eq: WF}, $V(\mathbf{r}) = \sum_i U(\mathbf{r} - \mathbf{r}_i)$ is the total impurity potential, and $U(\mathbf{r} - \mathbf{r}_i)$ is the potential due to a single impurity randomly located at position $\mathbf{r}_i$. The explicit form of $U(\mathbf{r}-\mathbf{r}_i)$ depends on the specific type of impurity potential considered. For simplicity, we assume that different types of impurity are uncorrelated, and their contributions to the total resistivity are additive ~\cite{Dugdale67pr}.

After solving the self-consistent equation, performing the ensemble average~\cite{AM07book}, and converting the summation over $k_x$ to the Landau level degeneracy via $(1/\mathcal{V}) \sum_{k_x} \rightarrow 1/(2\pi \ell_B^2)$ (see Sec. SII of~\cite{Supp}), the conductivity [Eq.~\eqref{Eq: sigma_AA}] is simplified to
\begin{equation}\label{Eq: sigma_theta}
\sigma_{xx}^{ql}=\frac{e^2 }{4 \pi h}\sum_{n=\pm 1}\sqrt{\frac{\vartheta_{n}}{\vartheta_0}}
\frac{1}{1+\vartheta_{n}/\varepsilon_{n}^2},
\end{equation}
where $\varepsilon_{n}$ is given by Eq.~\eqref{Eq: LL}. All aspects concerning impurity scattering are incorporated in the function $\vartheta_n$ ($n\in\{0,\pm1\}$), which is defined by 
\begin{equation}\label{Eq: theta}
\vartheta_{n}=\frac{n_i}{4^{|n|}}\int\frac{d\mathbf{q}}{(2\pi)^2}u_{\mathbf{q}}
u_{- \mathbf{q}} e^{- q^2 \ell_B^2 / 2}(\ell_Bq)^{2|n|},
\end{equation}
where $n_i$ is the impurity density, and $u_{\mathbf{q}}$ is the Fourier transform of the impurity potential, i.e., $u_{\mathbf{q}}=\int d\mathbf{r} e^{-i\mathbf{q}\cdot\mathbf{r}} U(\mathbf{r})$.

This expression [Eq.~\eqref{Eq: sigma_theta}] can be applicable to all types of impurity potentials, and is the central result of our work. In the quantum limit, the Fermi level only cuts the zeroth Landau level, which corresponds to a charge neutral point for 2D massless Dirac fermions. At this point, the Hall conductivity vanishes, and the magnetoresistivity simply reduces to $\rho_{xx}^{ql} = 1/\sigma_{xx}^{ql}$.

\begin{figure}[htbp]
\centering 
\includegraphics[width=0.48\textwidth]{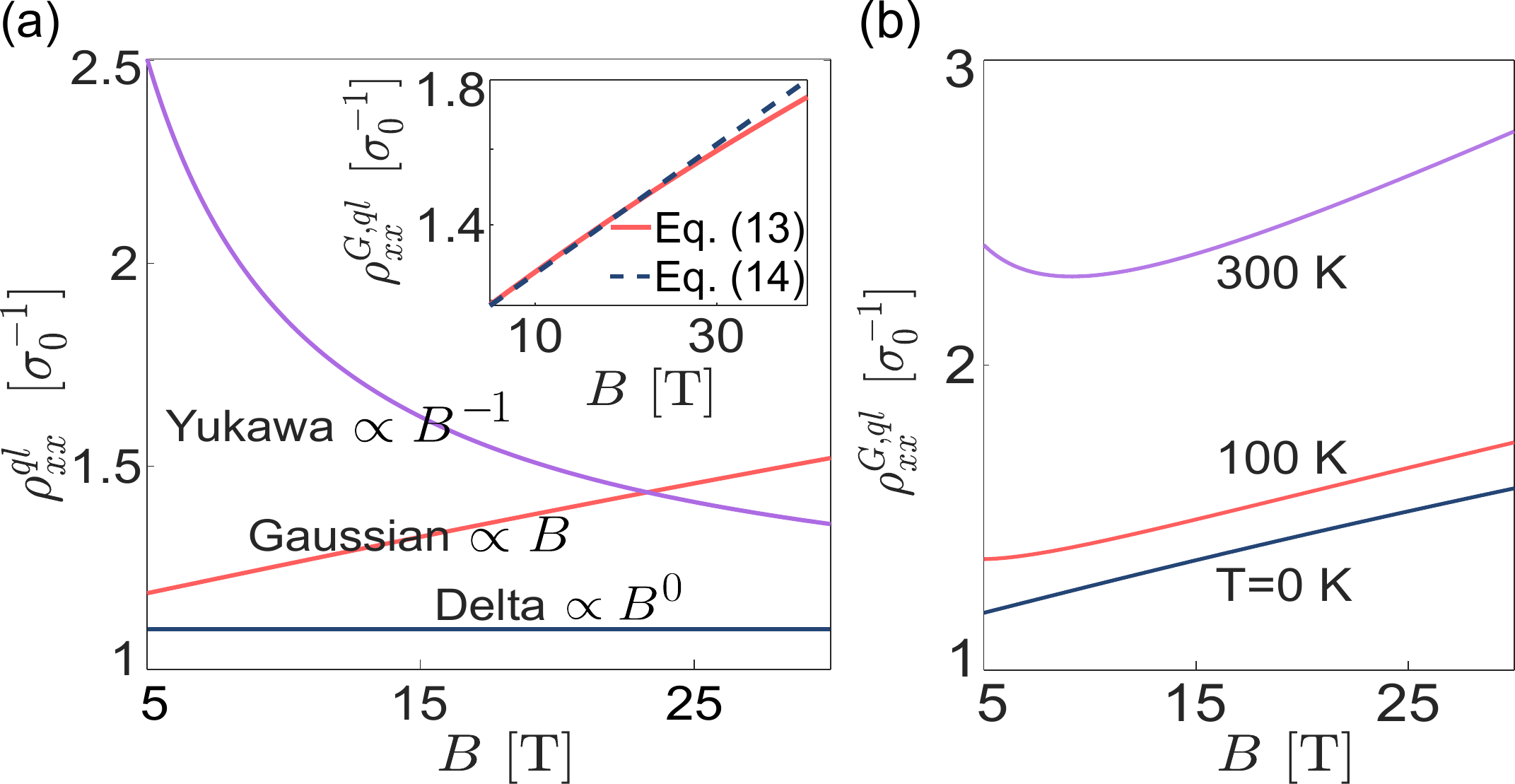}
\caption{(a) Comparison of calculated magnetoresistivities for different impurity potentials, as given by Eqs.~\eqref{Eq: rho_d}, \eqref{Eq: rho_G}, and \eqref{Eq: rho_Y}. The impurity strength is set to $\gamma = 0.1$ for both the $\delta$-function and Gaussian potentials. The decay length is taken $\lambda=4$ nm for the Gaussian potential. For the Yukawa potential, the impurity density is taken $n_i=0.1$ nm$^{-2}$. The other parameters are $\mathcal{C} \simeq 0.2$ and $g \simeq 1.22$ by adopting the experimental values $v=0.65$ eV$\cdot$nm~\cite{Geim05nature} and $\epsilon = 6.9\epsilon_0$~\cite{Massimo16prb}. The inset compares the original magnetoresistivity [Eq.~\eqref{Eq: rho_G}] with the approximated expression [Eq.~\eqref{Eq: rho_G2}] for the Gaussian potential. (b) Finite-temperature magnetoresistivity for the Gaussian potential at $T = 0$ K, 100 K, and 300 K.}
\label{Fig: compare} 
\end{figure}

\section{Magnetoresistivity for different impurity potentials}
\subsection{$\delta$-function potential}
%\textit{\textcolor{blue}{$\delta$-function potential.}--} 
For the simplest case, we first consider the point-like $\delta$-function potential (also known as the white noise). It takes the form
\begin{equation}
U(\mathbf{r}-\mathbf{r}_i)=u_0\delta(\mathbf{r}-\mathbf{r}_i)
\end{equation} 
with a constant $u_0$, and the corresponding Fourier component is $u_\mathbf{q}=u_0$. By substituting $u_\mathbf{q}$ into Eq.~\eqref{Eq: theta}, we first obtain $\vartheta_0=n_iu_0^2/(2\pi\ell_B^2)$ and $\vartheta_{\pm 1}= n_iu_0^2/(4\pi\ell_B^2)$. Then, inserting $\vartheta_{0}$ and $\vartheta_{\pm 1}$ into Eq.~\eqref{Eq: sigma_theta} and taking inverse of $\sigma_{xx}^{ql}$, the resistivity is given by
\begin{equation}\label{Eq: rho_d}
\rho_{xx}^{ql,\delta}=\sigma_0^{-1}\left(1+\gamma
\right),
\end{equation}
where $\sigma_0=e^2/(2\sqrt{2}\pi h)$ is the well-known minimal conductivity for 2D massless Dirac fermions~\cite{Ziegler97prb,Cserti07prb,Ziegler07prb,Dimi22prl}, and $\gamma = n_i u_{0}^2/(8\pi v^2)$ is a dimensionless quantity characterizing the impurity strength. Notably, this result is independent of the magnetic field. In the clean limit, $\gamma \rightarrow 0$, the resistivity $\rho_{xx}^{ql,\delta}$ simply reduces to $\sigma_0^{-1}$.

\subsection{Gaussian potential}
%\textit{\textcolor{blue}{Gaussian potential.}--} 
In contrast to the $\delta$-function potential, the Gaussian potential has a finite spatial spread and represents a long-range impurity. A typical form is given by
\begin{equation}
U(\mathbf{r}-\mathbf{r}_i)=u_0\left(\frac{1}{\sqrt{2\pi}\lambda}\right)^2e^{-\frac{|\mathbf{r}-\mathbf{r}_i|^2}{2\lambda^2}},
\end{equation}
where $\lambda$ is the decay length. Its corresponding Fourier component is given by $u_{\mathbf{q}} = u_0 e^{-q^2 \lambda^2 / 2}$. Accordingly, by using Eq.~\eqref{Eq: sigma_theta}, the longitudinal resistivity reads
\begin{equation}\label{Eq: rho_G}
\rho_{xx}^{ql,G}=\sigma_0^{-1}\sqrt{1+2\lambda^2/\ell_B^2}\left[1+\frac{\gamma}{(1+2\lambda^2/\ell_B^2)^2}\right],
\end{equation}
where the impurity strength $\gamma$ shares same definition with $\delta$-function impurity. Fig.~\ref{Fig: exp}(a) shows the behaviour of $\rho_{xx}^{ql, G}$ as a function of the dimensionless parameter $\lambda^2 / \ell_B^2$ for different values of $\gamma$.

In the limit $\lambda \rightarrow 0$, the expression reduces to the result for the $\delta$-function potential [Eq.~\eqref{Eq: rho_d}]. When the decay length $\lambda$ is smaller than the magnetic length $\ell_B$, the resistivity $\rho_{xx}^{ql,G}$ [Eq.~\eqref{Eq: rho_G}] can be approximately expanded as
\begin{equation}\label{Eq: rho_G2}
\rho_{xx}^{ql,G}\simeq\sigma_0^{-1}\left[1+\gamma+(1-3\gamma)\frac{\lambda^2eB}{\hbar}\right],
\end{equation}
where the magnetic length, defined as $\ell_B = \sqrt{\hbar / (eB)}$, has been substituted into the expression. As shown in the inset of Fig.~\ref{Fig: compare}(a), this expression [Eq.~\eqref{Eq: rho_G2}] exhibits excellent agreement with Eq.~\eqref{Eq: rho_G} across a broad range of magnetic fields. Notably, Eq.~\eqref{Eq: rho_G2} predicts a linear magnetoresistivity dependence on the magnetic field, with a slope given by $d\rho_{xx}^{ql,G}/dB=4\sqrt{2}\pi^2\lambda^2(1-3\gamma)/e$. Importantly, this linear magnetoresistance is a characteristic feature of 2D massless Dirac fermions with linear dispersion and does not occur in conventional 2D electron gases with quadratic dispersion (see Sec. SIII of~\cite{Supp}).

To validate our theoretical framework, we analyze two representative experimental datasets from Ref.~\cite{Geim23nature}, which has reported characteristic slopes of 1.2 and 7.3 k$\Omega/$T in the magnetoresistance measurements. As demonstrated in Fig.~\ref{Fig: exp}(b), our analytical result for $\rho_{xx}^{ql, G}$ [Eq.~\eqref{Eq: rho_G2}] successfully reproduces both the observed linear magnetoresistance behavior and, more importantly, quantitatively matches these distinct slopes. Additionally, as shown in the phase diagram of Fig.~\ref{Fig: exp}(c), our results accommodate a wide range of parameters that yield agreement with the experimental observations.

\subsection{Yukawa potential}
%\textit{\textcolor{blue}{Yukawa potential.}--} 
Finally, we consider the Yukawa potential (also known as the screened Coulomb potential), which describes the interaction between charge carriers and ionized impurities due to the screening effect in solids~\cite{Abrikosov98prb}. This potential takes the form
\begin{equation}
U(\mathbf{r}-\mathbf{r}_i)=\frac{e^2}{4\pi\epsilon|\mathbf{r}-\mathbf{r}_i|}e^{-\kappa|\mathbf{r}-\mathbf{r}_i|},
\end{equation}
where $\epsilon$ is the effective permittivity, and $\kappa$ is the inverse screening length (i.e., the Thomas–Fermi wavevector)~\cite{Ashcroft76book,AM07book}. Within the random phase approximation [Fig.~\ref{Fig: spectra}(e)]~\cite{Mahan90book,Abrikosov98prb,Qin20prl}, the inverse screening length is given by $\kappa = e^2 / (2\pi \epsilon \ell_B v)$ (see Sec. SII of~\cite{Supp}). 

The corresponding Fourier component takes the form $u_{\mathbf{q}} = e^2 / (2\epsilon \sqrt{q^2 + \kappa^2})$. Substituting $u_{\mathbf{q}}$ into Eq.~\eqref{Eq: sigma_theta}, the magnetoresistivity is obtained as
\begin{equation}\label{Eq: rho_Y}
\rho_{xx}^{ql,Y}=\frac{\sigma_0^{-1}}{\sqrt{1/(ge^{\mathcal{C}}) + \mathcal{C}}}\left[ 1 + \frac{\pi n_i\ell_B^2}{8}\mathcal{C} (1 + ge^{\mathcal{C}}\mathcal{C})\right],
\end{equation}
where $g$ denotes the incomplete gamma function evaluated at the dimensionless parameter $\mathcal{C}$, where $\mathcal{C}=e^4/(8\pi^2 \epsilon^2 v^2)$ is a $B$-independent dimensionless constant. 

In the limiting case $\mathcal{C} \rightarrow 0$, the expression again reduces to the result for the $\delta$-function potential. Since $\ell_B^2 = \hbar /(eB)$, the magnetoresistivity for the Yukawa potential follows $\rho_{xx}^{ql,Y} \propto B^{-1}$ as shown in Fig.~\ref{Fig: compare}(a). This behaviour contrasts sharply with the case of 3D Weyl metals, where the Yukawa potential gives rise to a linear magnetoresistivity~\cite{Abrikosov98prb}.

%\textit{\textcolor{blue}{Discussion and conclusion.}--}
\section{Discussion and conclusion}
Magnetoresistance constitutes a foundational tool for exploring the emergent properties of quantum materials. However, a consistent and detailed microscopic understanding of this phenomenon remains elusive across various platforms. 2D Dirac fermions in the quantum limit provide an analytically accessible platform for such investigations. Based on a concrete microscopic derivation, we obtain analytical expressions [Eqs.~\eqref{Eq: rho_d}, \eqref{Eq: rho_G}, and \eqref{Eq: rho_Y}] for magnetoresistivity arising from different impurity potentials. As shown in Fig.~\ref{Fig: compare}(a), these potentials yield qualitatively different magnetic-field dependencies. Although the  material-specific parameters dictate the detailed characteristics, the functional dependence on the magnetic field itself is universal. 
             
All the above results are obtained at zero temperature. Taking the Gaussian potential as a representative case, we evaluate the magnetoresistivity at finite temperatures by incorporating the Fermi–Dirac distribution, $\rho_{xx}^{ql,G}(T,B)=\int d\varepsilon\rho_{xx}^{ql,G}(0,B)[-\partial f_0/(\partial \varepsilon)]$, where $\rho_{xx}^{ql,G}(0,B)$ is the magnetoresistivity at zero temperature [Eq.~\eqref{Eq: rho_G}], $f_0$ is the Fermi-Dirac distribution. As shown in Fig.~\ref{Fig: compare}(b), even at the room temperature (300 K), the magnetoresistivity exhibits a robust linear dependence on the magnetic field over a wide range.

In conclusion, by combining the linear-response theory with the self‐consistent Born approximation, we obtain the analytical expressions for the longitudinal magnetoresistivity of 2D massless Dirac fermions in the quantum limit. We show that the $\delta$-function potential yields a field‐independent resistivity, which successfully recovers the minimal conductivity of 2D massless Dirac fermions. The Gaussian potential produces a pronounced linear magnetoresistivity, which quantitatively matches the experimental observations. Finally, the Yukawa potential gives rise to a $B^{-1}$-dependence, which is totally distinct from the known behaviour in the 3D system. These results bridge the theoretical gap of magnetoresistance for 2D massless Dirac fermions in the quantum limit.

\begin{acknowledgments}
This work is supported by the National Key R\&D Program of China (2022YFA1403700), the National Natural Science Foundation of China (12304074, 12350402, and 12525401), Guangdong Basic and Applied Basic Research Foundation (2023B0303000011), Guangdong Provincial Quantum Science Strategic Initiative (GDZX2201001 and GDZX2401001), the Science, Technology and Innovation Commission of Shenzhen Municipality (ZDSYS20190902092905285), High-level Special Funds (G03050K004), the New Cornerstone Science Foundation through the XPLORER PRIZE, and Center for Computational Science and Engineering of SUSTech.
\end{acknowledgments}

\bibliography{ref}
\end{document}

\title{Supplemental Material for\\
``Linear Magnetoresistance of 2D Massless Dirac Fermions in the Quantum Limit"}

\author{Xiao-Bin Qiang}
\email{These authors contributed equally to this work.}
\affiliation{State Key Laboratory of Quantum Functional Materials, Department of Physics, and Guangdong Basic Research Center of Excellence for Quantum Science, Southern University of Science and Technology (SUSTech), Shenzhen 518055, China}
\affiliation{Division of Physics and Applied Physics, School of Physical and Mathematical Sciences, Nanyang Technological University, 21 Nanyang Link, 637371, Singapore}

\author{Han-Yi Xu}
\email{These authors contributed equally to this work.}
\affiliation{State Key Laboratory of Quantum Functional Materials, Department of Physics, and Guangdong Basic Research Center of Excellence for Quantum Science, Southern University of Science and Technology (SUSTech), Shenzhen 518055, China}

\author{Ren-Jie Tong}
\email{These authors contributed equally to this work.}
\affiliation{State Key Laboratory of Quantum Functional Materials, Department of Physics, and Guangdong Basic Research Center of Excellence for Quantum Science, Southern University of Science and Technology (SUSTech), Shenzhen 518055, China}

\author{Shuai Li}
\affiliation{Hubei Engineering Research Center of Weak Magnetic-field Detection, Department of Physics, China Three Gorges University, Yichang 443002, China}

\author{Zi-Xuan Gao}
\affiliation{State Key Laboratory of Quantum Functional Materials, Department of Physics, and Guangdong Basic Research Center of Excellence for Quantum Science, Southern University of Science and Technology (SUSTech), Shenzhen 518055, China}

\author{Peng-Lu Zhao}
\email{Corresponding author: zhaopenglu@quantumsc.cn}
\affiliation{Quantum Science Center of Guangdong-Hong Kong-Macao Greater Bay Area (Guangdong), Shenzhen 518045, China}

\author{Hai-Zhou Lu}
\email{Corresponding author: luhz@sustech.edu.cn}
\affiliation{State Key Laboratory of Quantum Functional Materials, Department of Physics, and Guangdong Basic Research Center of Excellence for Quantum Science, Southern University of Science and Technology (SUSTech), Shenzhen 518055, China}
\affiliation{Quantum Science Center of Guangdong-Hong Kong-Macao Greater Bay Area (Guangdong), Shenzhen 518045, China}

\date{\today}

\begin{abstract}
This Supplemental Material provides detailed derivations and supporting calculations, which includes Refs. \cite{Neto09rmp,DasSarma11rmp,Mahan90book,Lu15prb,WangHW18prb,Abrikosov98prb,LiS23prb}.
\end{abstract}

\maketitle

\tableofcontents

\section{Landau Quantization for 2D Massless Dirac Fermions}
\subsection{2D Massless Dirac Fermions in Graphene}

The low-energy effective model of graphene at $\mathbf{K}$-point is characterized by 2D massless Dirac model, which reads~\cite{Neto09rmp,DasSarma11rmp}
\begin{equation}
\mathcal{H}_{\mathbf{K}}=v(k_x\sigma_x+k_y\sigma_y),
\end{equation}
while at $\mathbf{K}'$-point
\begin{equation}
\mathcal{H}_{\mathbf{K}'}=v(k_x\sigma_x-k_y\sigma_y),
\end{equation}
where $v$ is the model parameter, $\mathbf{k}$ is the wave vector, and $\sigma_i$'s are Pauli matrices. These two models can be connected by a unitary transformation $\sigma_x\mathcal{H}_{\mathbf{K}}\sigma_x^\dagger=\mathcal{H}_{\mathbf{K}'}$.

\subsection{Landau Quantization at $\mathbf{K}$-point}

To calculate the Landau level, we first consider the effective model at $\mathbf{K}$-point, which can be rewritten as 
\begin{equation}
\begin{aligned}
\mathcal{H}_{\mathbf{K}}&=v(k_x\sigma_x+k_y\sigma_y)\\
&=\begin{bmatrix}
0 & v(k_x-ik_y)\\
v(k_x+ik_y) & 0
\end{bmatrix}\\
&=\begin{bmatrix}
0 & vk_-\\
vk_+ & 0
\end{bmatrix},\\
\end{aligned}
\end{equation}
where $k_{\pm}$ is defined as $k_{\pm}= k_x\pm ik_y$. We adopt the Landau gauge $\mathbf{A}=(-yB,0,0)$, and the Peierls substitution $k_x\rightarrow k_x-eBy/\hbar$. Then, $k_\pm$ become
\begin{equation}
\begin{aligned}
k_+&\rightarrow k_x-\frac{eBy}{\hbar}+i(-i\partial_y)\\
&=-\frac{1}{\ell_B}\left(\frac{y-y_0}{\ell_B}-\ell_B\partial_y\right)\\
&=\frac{\sqrt{2}}{\ell_B}a^\dagger,\\
k_-&\rightarrow k_x-\frac{eBy}{\hbar}-i(-i\partial_y)\\
&=-\frac{1}{\ell_B}\left(\frac{y-y_0}{\ell_B}+\ell_B\partial_y\right)\\
&=\frac{\sqrt{2}}{\ell_B}a,
\end{aligned}
\end{equation}
where $\ell_B=\sqrt{\hbar/(eB)}$ is the magnetic length, $y_0=\ell_B^2k_x$ is the guiding center, and the ladder operators are defined as
\begin{equation}
\left\{
\begin{aligned}
a^\dagger&=-\frac{1}{\sqrt{2}}\left[\frac{y-y_0}{\ell_B}-\ell_B\partial_y\right],\\
a&=-\frac{1}{\sqrt{2}}\left[\frac{y-y_0}{\ell_B}+\ell_B\partial_y\right].\\
\end{aligned}\right.
\end{equation}

Then, the the Hamiltonian becomes 
\begin{equation}
\begin{aligned}
\mathcal{H}_{\mathbf{K}}&=\begin{bmatrix}
0 &  (\sqrt{2}v/\ell_B) a\\
(\sqrt{2}v/\ell_B) a^\dagger & 0
\end{bmatrix}
=\begin{bmatrix}
0 & \beta a\\
\beta a^\dagger & 0
\end{bmatrix}\\
\end{aligned},
\end{equation}
where we have defined $\beta = \sqrt{2}v/\ell_B$ for the simplicity of notation. By using the trial wave function $[c_1|n-1\rangle, c_2|n\rangle]^T$, the eigenequation now reads
\begin{equation}
\begin{aligned}
\begin{bmatrix}
0 & \beta a\\
\beta a^\dagger & 0
\end{bmatrix}
\begin{bmatrix}
c_1|n-1\rangle\\
c_2|n\rangle
\end{bmatrix}=\varepsilon\begin{bmatrix}
c_1|n-1\rangle\\
c_2|n\rangle
\end{bmatrix},\qquad(n=1,2\cdots),\\
\end{aligned}
\end{equation}
where $|n\rangle$ is the eigenstate of the harmonic oscillator. Then, the eigenequation can be rewritten as 
\begin{equation}
\begin{aligned}
\left\{ \begin{aligned}
c_2\beta a|n\rangle=c_1\varepsilon|n-1\rangle,\\
c_1\beta a^\dagger|n-1\rangle=c_2\varepsilon|n\rangle.
\end{aligned}
\right.\\
\end{aligned}
\end{equation}
Recalling the properties of ladder operators $a|n\rangle=\sqrt{n}|n-1\rangle$ and $a^\dagger|n\rangle=\sqrt{n+1}|n+1\rangle$, we have
\begin{equation}
\begin{aligned}
\left\{ \begin{aligned}
&c_2\beta \sqrt{n}|n-1\rangle=c_1\varepsilon|n-1\rangle,\\
&c_1\beta\sqrt{n}|n\rangle=c_2\varepsilon|n\rangle.
\end{aligned}
\right.
\end{aligned}
\end{equation}
Now, we obtain a new eigenequation
\begin{equation}
\begin{aligned}
\begin{bmatrix}
0 & \beta \sqrt{n}\\
\beta \sqrt{n} & 0
\end{bmatrix}
\begin{bmatrix}
c_1 \\
c_2
\end{bmatrix}=\varepsilon\begin{bmatrix}
c_1 \\
c_2
\end{bmatrix}\\
\end{aligned}.
\end{equation}
Therefore, the solutions read
\begin{equation}\label{Eq: En_K}
\begin{aligned}
\varepsilon_{\pm n}=\pm\sqrt{\beta^2n}=\pm\frac{1}{\ell_B}\sqrt{2v^2n},\qquad(n=1,2\cdots).
\end{aligned}
\end{equation}

For the zeroth Landau level, we use the trial wave function $[ 0, |0\rangle]^T$, the eigenfunction reads
\begin{equation}
\begin{aligned}
\begin{bmatrix}
0 & \beta a\\
\beta a^\dagger & 0
\end{bmatrix}
\begin{bmatrix}
0\\
|0\rangle
\end{bmatrix}&=\varepsilon\begin{bmatrix}
0\\
|0\rangle
\end{bmatrix}.\\
\end{aligned}
\end{equation}
The solution is 
\begin{equation}
\begin{aligned}
\varepsilon_0=0.
\end{aligned}
\end{equation}
These eigenenergies represent a series of Landau levels ($n$ as level index). For $n\ge 1$, the eigenstates are
\begin{equation}
\begin{aligned}
|k_x,+n\rangle&=\frac{1}{\sqrt{2}}
\begin{bmatrix}
|n-1\rangle\\
|n\rangle
\end{bmatrix}|k_x\rangle,\\
|k_x,-n\rangle&=\frac{1}{\sqrt{2}}
\begin{bmatrix}
|n-1\rangle\\
- |n\rangle
\end{bmatrix}|k_x\rangle.
\end{aligned}
\end{equation}
For $n=0$, the eigenstate is
\begin{equation}
|k_x,0\rangle=
\begin{bmatrix}
0 \\ |0\rangle  
\end{bmatrix}
|k_x\rangle.
\end{equation}

Additionally, the wave functions can be found as
\begin{equation}
\begin{aligned}
\psi_{k_x ,+n}(\mathbf{r})&=\langle\mathbf{r}|k_x ,+n\rangle=\frac{e^{ik_x x}}{\sqrt{2L_x}}  
\begin{bmatrix}
\phi_{k_x,n-1}(y) \\
\phi_{k_x,n}(y) \\
\end{bmatrix},\\
\psi_{k_x ,-n}(\mathbf{r})&=\langle\mathbf{r}|k_x ,-n\rangle=\frac{e^{ik_x x}}{\sqrt{2L_x}}  
\begin{bmatrix}
\phi_{k_x,n-1}(y)\\
- \phi_{k_x,n}(y)\\
\end{bmatrix},\\
\psi_{k_x,0}(\mathbf{r})&=\langle\mathbf{r}|k_x,0\rangle=\frac{e^{ik_x x}}{\sqrt{L_x}}  
\begin{bmatrix}
0 \\
\phi_{k_x,0}(y) \\
\end{bmatrix}\\
\end{aligned}
\end{equation}
with
\begin{equation}
\phi_{k_x,n}(y)=\frac{1}{\sqrt{n!2^n\sqrt{\pi}\ell_B}}e^{-\zeta^2/2}H_n(\zeta),
\end{equation}
where $\zeta=(y-y_0)/\ell_B$, $y_0=\ell_B^2k_x$ is the guiding center, and $H_n$ is the Hermite polynomial.

\subsection{Landau Quantization at $\mathbf{K}'$-point}

We now consider the effective model at $\mathbf{K}'$-point, which can be rewritten as 
\begin{equation}
\begin{aligned}
\mathcal{H}_{\mathbf{K}'}&=v(k_x\sigma_x-k_y\sigma_y)\\
&=\begin{bmatrix}
0 & v(k_x+ik_y)\\
v(k_x-ik_y) & 0
\end{bmatrix}\\
&=\begin{bmatrix}
0 & vk_+\\
vk_- & 0
\end{bmatrix},\\
\end{aligned}
\end{equation}
where $k_{\pm}$ is defined as $k_{\pm}= k_x\pm ik_y$. Under the Landau gauge $\mathbf{A}=(-yB,0,0)$, and applying the Peierls substitution $k_x\rightarrow k_x-eBy/\hbar$, $k_\pm$ read 
\begin{equation}
\begin{aligned}
k_+\rightarrow \frac{\sqrt{2}}{\ell_B}a^\dagger,\quad k_-\rightarrow \frac{\sqrt{2}}{\ell_B}a.
\end{aligned}
\end{equation}

Then, the the Hamiltonian becomes 
\begin{equation}
\begin{aligned}
\mathcal{H}_{\mathbf{K}'}&=\begin{bmatrix}
0 &  (\sqrt{2}v/\ell_B) a^\dagger\\
(\sqrt{2}v/\ell_B) a & 0
\end{bmatrix}
=\begin{bmatrix}
0 & \beta a^\dagger\\
\beta a & 0
\end{bmatrix}\\
\end{aligned},
\end{equation}
where $\beta = \sqrt{2}v/\ell_B$. By using the trial wave function $[c_1|n\rangle, c_2|n-1\rangle]^T$, the eigenequation now reads
\begin{equation}
\begin{aligned}
\begin{bmatrix}
0 & \beta a^\dagger\\
\beta a & 0
\end{bmatrix}
\begin{bmatrix}
c_1|n\rangle\\
c_2|n-1\rangle
\end{bmatrix}=\varepsilon\begin{bmatrix}
c_1|n\rangle\\
c_2|n-1\rangle
\end{bmatrix},\qquad(n=1,2\cdots),\\
\end{aligned}
\end{equation}
where $|n\rangle$ is the eigenstate of the harmonic oscillator. Then, the eigenequation can be rewritten as 
\begin{equation}
\begin{aligned}
\left\{\begin{aligned}
&c_2\beta a^\dagger|n-1\rangle=c_1\varepsilon|n\rangle,\\
&c_1\beta a|n\rangle=c_2\varepsilon|n-1\rangle.
\end{aligned}
\right.\\
\end{aligned}
\end{equation}
Recalling the properties of ladder operators $a|n\rangle=\sqrt{n}|n-1\rangle$ and $a^\dagger|n\rangle=\sqrt{n+1}|n+1\rangle$, we have
\begin{equation}
\begin{aligned}
\left\{ \begin{aligned}
&c_2\beta \sqrt{n}|n\rangle=c_1\varepsilon|n\rangle,\\
&c_1\beta\sqrt{n}|n-1\rangle=c_2\varepsilon|n-1\rangle.
\end{aligned}
\right.
\end{aligned}
\end{equation}
Now, we obtain a new eigenequation
\begin{equation}
\begin{aligned}
\begin{bmatrix}
0 & \beta \sqrt{n}\\
\beta \sqrt{n} & 0
\end{bmatrix}
\begin{bmatrix}
c_1 \\
c_2
\end{bmatrix}=\varepsilon\begin{bmatrix}
c_1 \\
c_2
\end{bmatrix}\\
\end{aligned}.
\end{equation}
Therefore, the solutions read
\begin{equation}
\begin{aligned}
\varepsilon_{\pm n}=\pm\sqrt{\beta^2n}=\pm\frac{1}{\ell_B}\sqrt{2v^2n},\qquad(n=1,2\cdots).
\end{aligned}
\end{equation}
This result exactly equals to Eq.~\eqref{Eq: En_K}.

For the zeroth Landau level, we use the trial wave function $[|0\rangle,0]^T$, the eigenfunction reads
\begin{equation}
\begin{aligned}
\begin{bmatrix}
0 & \beta a^\dagger\\
\beta a & 0
\end{bmatrix}
\begin{bmatrix}
|0\rangle\\
0
\end{bmatrix}&=\varepsilon\begin{bmatrix}
|0\rangle\\
0
\end{bmatrix}.\\
\end{aligned}
\end{equation}
The solution is 
\begin{equation}
\begin{aligned}
\varepsilon_0=0.
\end{aligned}
\end{equation}
These eigenenergies represent a series of Landau levels ($n$ as level index). For $n\ge 1$, the eigenstates are
\begin{equation}
\begin{aligned}
|k_x,+n\rangle&=\frac{1}{\sqrt{2}}
\begin{bmatrix}
|n\rangle\\
|n-1\rangle
\end{bmatrix}|k_x\rangle,\\
|k_x,-n\rangle&=\frac{1}{\sqrt{2}}
\begin{bmatrix}
|n\rangle\\
-|n-1\rangle
\end{bmatrix}|k_x\rangle.
\end{aligned}
\end{equation}
For $n=0$, the eigenstate is
\begin{equation}
|k_x,0\rangle=
\begin{bmatrix}
|0\rangle \\ 0  
\end{bmatrix}
|k_x\rangle.
\end{equation}

Additionally, the wave functions can be found as
\begin{equation}
\begin{aligned}
\psi_{k_x ,+n}(\mathbf{r})&=\langle\mathbf{r}|k_x ,+n\rangle=\frac{e^{ik_x x}}{\sqrt{2L_x}}  
\begin{bmatrix}
\phi_{k_x,n}(y) \\
\phi_{k_x,n-1}(y) \\
\end{bmatrix},\\
\psi_{k_x ,-n}(\mathbf{r})&=\langle\mathbf{r}|k_x ,-n\rangle=\frac{e^{ik_x x}}{\sqrt{2L_x}}  
\begin{bmatrix}
\phi_{k_x,n}(y)\\
- \phi_{k_x,n-1}(y)\\
\end{bmatrix},\\
\psi_{k_x,0}(\mathbf{r})&=\langle\mathbf{r}|k_x,0\rangle=\frac{e^{ik_x x}}{\sqrt{L_x}}  
\begin{bmatrix}
\phi_{k_x,0}(y) \\
0
\end{bmatrix}\\
\end{aligned}
\end{equation}
with
\begin{equation}
\phi_{k_x,n}(y)=\frac{1}{\sqrt{n!2^n\sqrt{\pi}\ell_B}}e^{-\zeta^2/2}H_n(\zeta),
\end{equation}
where $\zeta=(y-y_0)/\ell_B$, $y_0=\ell_B^2k_x$ is the guiding center, and $H_n$ is the Hermite polynomial.

\subsection{Degeneracy of Landau Levels}

The eigenenergies of 2D massless Dirac fermions without magnetic field are
\begin{equation} 
\varepsilon_\pm=\pm v\sqrt{k_x^2+k_y^2}=\pm vk.
\end{equation}
The density of states can be calculated as
\begin{equation}
\begin{aligned}
\int \rho(\varepsilon)d\varepsilon&=\int\frac{d^2 \mathbf{k}}{(2\pi)^2}=\int\frac{d\varphi kdk}{(2\pi)^2}=\int\frac{kdk}{2\pi},\\
\Rightarrow\quad& \rho(\varepsilon)d\varepsilon=\rho(\varepsilon)vdk=\frac{kdk}{2\pi},\\
\Rightarrow\quad& \rho(\varepsilon)=\frac{k}{2\pi v}=\frac{\varepsilon}{2\pi v^2}.
\end{aligned}
\end{equation}
Now, the degeneracy of Landau levels is
\begin{equation}
\begin{aligned}
N_{L} & = \int_{\varepsilon_n}^{\varepsilon_{n+1}} \rho(\varepsilon) d \varepsilon = \int_{\varepsilon_n}^{\varepsilon_{n+1}}\frac{\varepsilon}{2\pi v^2}d\varepsilon=\frac{1}{2\pi\ell_B^2}.
\end{aligned}
\end{equation}

\section{Calculation for Magnetoresistivity in the Quantum Limit}

In the quantum limit, the Fermi level cuts the lowest Landau level. We assume that the magnetic field along the $\hat{z}$ direction. Recalling the
Kubo formula for magnetotransport, the conductivity is
\begin{equation}
\sigma_{xx} = \frac{e^2 \hbar}{4 \pi \mathcal{V}}\sum_{l, l'} |v_{l,
l'}^x |^2 \mathcal{A}_l (\varepsilon_F) \mathcal{A}_{l'} (\varepsilon_F),
\end{equation}
where the summation over $l, l'$ represents
\begin{equation}
\sum_{l, l'}\rightarrow \sum_{k_x, k_x'}\sum_{n, n'} .
\end{equation}
Thus, in the quantum limit ($n = 0$), the conductivity can be
written as
\begin{equation}
\begin{aligned}
\sigma_{xx}^{ql} & = \frac{e^2 \hbar}{4 \pi \mathcal{V}}\sum_{k_x,
k_x'}\sum_{n'} |v_{k_x, 0 ; k_x', n'}^x |^2 \mathcal{A}_{k_x, 0}
(\varepsilon_F) \mathcal{A}_{k_x', n'} (\varepsilon_F) .
\end{aligned}
\end{equation}
To be continue, we firstly need the matrix elements for the velocity operator
in the Landau basis. The velocity operator is given by
\begin{equation}
\begin{aligned}
v_x & = \frac{1}{i \hbar}  [x, \mathcal{H} (\mathbf{k} + e \mathbf{A})]
= \frac{1}{i \hbar}\left[ x, v \big( k_x - \frac{e}{\hbar} By \big)
\sigma_x + vk_y \sigma_y \right] = \frac{v}{\hbar}\sigma_x,
\end{aligned}
\end{equation}
corresponding matrix elements read
\begin{equation}
\left\{
\begin{aligned}
& |v_{k_x, 0 ; k_x', 0}^x |^2 = | \langle k_x, 0| v_x |k_x', 0 \rangle
|^2 = 0,\\
& |v_{k_x, 0 ; k_x', + n'}^x |^2 = | \langle k_x, 0| v_x |k_x', + n'
\rangle |^2 = \frac{v^2}{2 \hbar^2}\delta_{0, + n' - 1}\delta_{k_x,
k_x'},\\
& |v_{k_x, 0 ; k_x', - n'}^x |^2 = | \langle k_x, 0| v_x |k_x', - n'
\rangle |^2 = \frac{v^2}{2 \hbar^2}\delta_{0, - n' - 1}\delta_{k_x,
k_x'}.
\end{aligned}\right.
\end{equation}
Notably, the velocity matrix elements are identical for the Landau bases at both $\mathbf{K}$ and $\mathbf{K}'$ valleys.

Thus, we have
\begin{equation}
\begin{aligned}
\sigma_{xx}^{ql} & = \frac{e^2 \hbar}{4 \pi \mathcal{V}}\sum_{k_x,
k_x'}\sum_{n'} |v_{k_x, 0 ; k_x', n'}^x |^2 \mathcal{A}_{k_x, 0}
(\varepsilon_F) \mathcal{A}_{k_x', n'} (\varepsilon_F)\\
& = \frac{e^2 v^2}{8 \pi \hbar \mathcal{V}}\sum_{k_x}\left[ \sum_{n'}
\delta_{0, + n' - 1}\mathcal{A}_{k_x, 0}
(\varepsilon_F)\mathcal{A}_{k_x, n'} (\varepsilon_F) + \sum_{n'}
\delta_{0, - n' - 1}\mathcal{A}_{k_x, 0}
(\varepsilon_F)\mathcal{A}_{k_x, n'} (\varepsilon_F) \right]\\
& = \frac{e^2 v^2}{8 \pi \hbar \mathcal{V}}\sum_{k_x}
[\mathcal{A}_{k_x, 0} (\varepsilon_F)\mathcal{A}_{k_x, + 1}
(\varepsilon_F) +\mathcal{A}_{k_x, 0} (\varepsilon_F)\mathcal{A}_{k_x, -1} (\varepsilon_F)]\\
& = \sigma_{xx, + 1}^{ql} + \sigma_{xx, - 1}^{ql}
\end{aligned}
\end{equation}
with
\begin{equation}\label{Eq: sigmaxx_Dirac}
\sigma_{xx, \pm 1}^{ql} = \frac{e^2 v^2}{8 \pi \hbar \mathcal{V}} 
\sum_{k_x}\mathcal{A}_{k_x, 0} (\varepsilon_F) \mathcal{A}_{k_x, \pm 1}
(\varepsilon_F) .
\end{equation}
Here, the spectral functions read
\begin{equation}
\left\{
\begin{aligned}
&\mathcal{A}_{k_x, 0} (\varepsilon_F) = \frac{2 \Delta_{k_x,
0}}{(\varepsilon_F - \varepsilon_0)^2 + \Delta_{k_x, 0}^2},\\
&\mathcal{A}_{k_x, \pm 1} (\varepsilon_F) = \frac{2 \Delta_{k_x, \pm
1}}{(\varepsilon_F - \varepsilon_{\pm 1})^2 + \Delta_{k_x, \pm 1}^2},
\end{aligned}\right.
\end{equation}
and the line-width function is given by
\begin{equation}
\Delta_{k_x, n} = - \text{Im} [\Sigma_{k_x, n}^r],
\end{equation}
where $\Sigma_{k_x, n}^r$ is the selfenergy due to the impurity scattering.

\subsection{Selfenergy $\Sigma_0^r$}

By using self-consistent Born approximation [Fig. 2(d) in the article], the selfenergy for $n = 0$
Landau level is
\begin{equation}
\begin{aligned}
\Sigma_0^r &= \sum_{k_x', n'}\frac{| \langle k_x', n' |V|k_x, 0 \rangle
|^2}{\varepsilon_F - \varepsilon_{n'} - \Sigma_{n'}^r},
\end{aligned}
\end{equation}
where $V= \sum_i U(\mathbf{r} - \mathbf{r}_i)$ is the total impurity potential, and $U(\mathbf{r} - \mathbf{r}_i)$ is the potential due to a single impurity randomly located at position $\mathbf{r}_i$. Due to the condition of quantum limit, only the $n' = 0$ term is dominant. The scattering matrix elements in spatial
representation is given by
\begin{equation}
\begin{aligned}
\langle k_x', 0| V|k_x, 0 \rangle & = \int d \mathbf{r}\int d
\mathbf{r}' \langle k_x', 0| \mathbf{r}\rangle \langle \mathbf{r} |V|
\mathbf{r}' \rangle \langle \mathbf{r}' |k_x, 0 \rangle\\
& = \int d \mathbf{r}\int d \mathbf{r}' \langle k_x', 0| \mathbf{r}
\rangle V (\mathbf{r}) \delta (\mathbf{r} - \mathbf{r}')  \langle
\mathbf{r}' |k_x, 0 \rangle\\
& = \int d \mathbf{r}\langle k_x', 0| \mathbf{r}\rangle V (\mathbf{r})
\langle \mathbf{r} |k_x, 0 \rangle\\
& = \int d \mathbf{r}\psi_{k_x', 0}^{\dagger} (\mathbf{r}) V
(\mathbf{r}) \psi_{k_x, 0} (\mathbf{r}),
\end{aligned}
\end{equation}
where $\langle \mathbf{r} |V| \mathbf{r}' \rangle = V (\mathbf{r}) \delta
(\mathbf{r} - \mathbf{r}')$, and the wave function of the zeroth Landau level
is
\begin{equation}
\psi_{k_x, 0} (\mathbf{r}) = \frac{e^{ik_x x}}{\sqrt{L_x}}\begin{bmatrix}
0\\
\phi_{k_x, 0} (y)
\end{bmatrix},
\end{equation}
where
\begin{equation}
\phi_{k_x, 0} (y) = \frac{1}{\sqrt{\sqrt{\pi}\ell_B}} e^{- \zeta^2 / 2}
\end{equation}
with $\zeta = y / \ell_B - \ell_B k_x$. Recalling the ensemble average
\begin{equation}
\langle V (\mathbf{r}) V (\mathbf{r}') \rangle_{dis} = n_i  \int [d
\mathbf{q}] e^{i \mathbf{q}\cdot (\mathbf{r} - \mathbf{r}')}
u_{\mathbf{q}} u_{- \mathbf{q}},
\end{equation}
we can obtain that
\begin{equation}
\begin{aligned}
| \langle k_x', 0| V|k_x, 0 \rangle |^2 & = \frac{n_i}{L_x^2}\int [d
\mathbf{q}] u_{\mathbf{q}} u_{- \mathbf{q}}\int d \mathbf{r}\int d
\mathbf{r}' e^{i \mathbf{q}\cdot (\mathbf{r} - \mathbf{r}')} e^{i (k_x -
k_x') x}\\
& \quad \times e^{- i (k_x - k_x') x'}\phi_{k_x, 0} (y) \phi_{k_x', 0}
(y) \phi_{k_x, 0} (y') \phi_{k_x', 0} (y')\\
& = n_i  \int [d \mathbf{q}] u_{\mathbf{q}} u_{- \mathbf{q}}\int dy
\int dy' e^{iq_y  (y - y')}\phi_{k_x, 0} (y) \phi_{k_x', 0} (y)\\
& \quad \times \phi_{k_x, 0} (y') \phi_{k_x', 0} (y') \delta_{q_x + k_x
- k_x'}\delta_{- q_x - k_x + k_x'}\\
& = n_i  \int [d \mathbf{q}] u_{\mathbf{q}} u_{- \mathbf{q}} e^{- [(k_x
- k_x')^2 + q_y^2] \ell_B^2 / 2}\delta_{q_x + k_x - k_x'},
\end{aligned}
\end{equation}
where we have used
\begin{equation}
\begin{aligned}
\int dye^{\pm iq_y y}\phi_{k_x, 0} (y) \phi_{k_x', 0} (y) & =
\frac{1}{\ell_B  \sqrt{\pi}}\int dye^{\pm iq_y y} e^{- (y / \ell_B -
\ell_B k_x)^2 / 2} e^{- (y / \ell_B - \ell_B k_x')^2 / 2}\\
& = e^{- [(k_x - k_x')^2 + q_y^2] \ell_B^2 / 4 \pm iq_y  (k_x + k_x')
\ell_B^2 / 2} .
\end{aligned}
\end{equation}
Thus, the final result of the selfenergy for $n = 0$ Landau level is
\begin{equation}
\begin{aligned}
\Sigma_0^r & \simeq \sum_{k_x'}\frac{| \langle k_x', 0| V|k_x, 0 \rangle
|^2}{\varepsilon_F - \varepsilon_0 - \Sigma_0^r}\\
& = \frac{n_i }{\varepsilon_F - \Sigma_0^r}\sum_{k_x'}\int [d
\mathbf{q}] u_{\mathbf{q}} u_{- \mathbf{q}} e^{- [(k_x - k_x')^2 + q_y^2]
\ell_B^2 / 2}\delta_{q_x + k_x - k_x'}\\
& = \frac{n_i }{\varepsilon_F - \Sigma_0^r}\int [d \mathbf{q}]
u_{\mathbf{q}} u_{- \mathbf{q}} e^{- q^2 \ell_B^2 / 2},
\end{aligned}
\end{equation}
where $\varepsilon_0 = 0$ and $q^2 = q_x^2 + q_y^2$. This is a general
expression for all impurity potentials.

\subsection{Selfenergies $\Sigma_{\pm 1}^r$}

The selfenergy for $n=\pm 1$ Landau level is given by
\begin{equation}
\begin{aligned}
\Sigma_{\pm 1}^r & = \sum_{k_x', n'}\frac{| \langle k_x', n' |V|k_x, \pm
1 \rangle |^2}{\varepsilon_F - \varepsilon_{n'} - \Sigma_{\pm 1}^r} .
\end{aligned}
\end{equation}
Due to the condition of quantum limit, only the $n' = 0$ term is dominant. The
scattering matrix element reads
\begin{equation}
\langle k_x', 0| V|k_x, \pm 1 \rangle = \int d \mathbf{r}
\psi_{k_x', 0}^{\dagger} (\mathbf{r}) V (\mathbf{r}) \psi_{k_x, \pm 1}
(\mathbf{r}),
\end{equation}
and the wave functions are
\begin{equation}
\left\{
\begin{aligned}
&\psi_{k_x, 0} (\mathbf{r}) = \frac{e^{ik_x x}}{\sqrt{L_x}}\begin{bmatrix}
0\\
\phi_{k_x, 0} (y)
\end{bmatrix},\\
&\psi_{k_x, \pm 1} (\mathbf{r})  = \frac{e^{ik_x x}}{\sqrt{2 L_x}}\begin{bmatrix}
\phi_{k_x, 0} (y)\\
\pm \phi_{k_x, 1} (y)
\end{bmatrix},
\end{aligned}\right.
\end{equation}
where
\begin{equation}
\left\{
\begin{aligned}
\phi_{k_x, 0} (y) & = \frac{1}{\sqrt{\sqrt{\pi}\ell_B}} e^{- \zeta^2 /
2},\\
\phi_{k_x, 1} (y) & = \frac{\sqrt{2}\zeta}{\sqrt{\sqrt{\pi}\ell_B}}
e^{- \zeta^2 / 2}
\end{aligned}\right.
\end{equation}
with $\zeta = y / \ell_B - \ell_B k_x$. Recalling the ensemble average
\begin{equation}
\langle V (\mathbf{r}) V (\mathbf{r}') \rangle_{dis} = n_i  \int [d
\mathbf{q}] e^{i \mathbf{q}\cdot (\mathbf{r} - \mathbf{r}')}
u_{\mathbf{q}} u_{- \mathbf{q}},
\end{equation}
we can obtain that
\begin{equation}
\begin{aligned}
| \langle k_x', 0| V|k_x, \pm 1 \rangle |^2 & = \frac{n_i}{2 L_x^2}\int
[d \mathbf{q}] u_{\mathbf{q}} u_{- \mathbf{q}}\int d \mathbf{r}\int d
\mathbf{r}' e^{i \mathbf{q}\cdot (\mathbf{r} - \mathbf{r}')} e^{i(k_x-k_x')x}\\
& \quad \times e^{- i (k_x - k_x') x'}\phi_{k_x, 0} (y) \phi_{k_x', 1}
(y) \phi_{k_x, 0} (y') \phi_{k_x', 1} (y')\\
& = \frac{n_i}{2}\int [d \mathbf{q}] u_{\mathbf{q}} u_{- \mathbf{q}} 
\int dy \int dy' e^{iq_y  (y - y')}\phi_{k_x, 0} (y) \phi_{k_x', 1}
(y)\\
& \quad \times \phi_{k_x, 0} (y') \phi_{k_x', 1} (y') \delta_{q_x + k_x
- k_x'}\delta_{- q_x - k_x + k_x'}\\
& = \frac{n_i}{4}\int [d \mathbf{q}] u_{\mathbf{q}} u_{- \mathbf{q}}
e^{- [(k_x - k_x')^2 + q_y^2] \ell_B^2 / 2}\ell_B^2  [(k_x - k_x')^2 +
q_y^2] \delta_{q_x + k_x - k_x'},
\end{aligned}
\end{equation}
where we have used
\begin{equation}
\begin{aligned}
\int dye^{\pm iq_y y}\phi_{k_x, 0} (y) \phi_{k_x', 1} (y) & =
\frac{\sqrt{2}}{\ell_B  \sqrt{\pi}}\int dy (y / \ell_B - \ell_B k_x)
e^{\pm iq_y y} e^{- (y / \ell_B - \ell_B k_x)^2 / 2} e^{- (y / \ell_B -
\ell_B k_x')^2 / 2}\\
& = \frac{\ell_B}{\sqrt{2}} e^{- [(k_x - k_x')^2 + q_y^2] \ell_B^2 / 4
\pm iq_y  (k_x + k_x') \ell_B^2 / 2}  (- k_x + k_x' \pm iq_y) .
\end{aligned}
\end{equation}
Thus, the final result of the self-energy for $n=\pm 1$ Landau level is
\begin{equation}
\begin{aligned}
\Sigma_{\pm 1}^r &\simeq \frac{| \langle k_x', n' |V|k_x, \pm 1 \rangle
|^2}{\varepsilon_F - \varepsilon_0 - \Sigma_{\pm 1}^r}\\
& = \frac{1}{\varepsilon_F - \Sigma_{\pm 1}^r}\frac{n_i}{4}
\sum_{k_x'}\int [d \mathbf{q}] u_{\mathbf{q}} u_{- \mathbf{q}} e^{-
[(k_x - k_x')^2 + q_y^2] \ell_B^2 / 2}\ell_B^2  [(k_x - k_x')^2 + q_y^2]
\delta_{q_x + k_x - k_x'}\\
& = \frac{1}{\varepsilon_F - \Sigma_{\pm 1}^r}\frac{n_i}{4}\int [d
\mathbf{q}] u_{\mathbf{q}} u_{- \mathbf{q}} e^{- q^2 \ell_B^2 / 2}
\ell_B^2 q^2,
\end{aligned}
\end{equation}
where $\varepsilon_0 = 0$ and $q^2 = q_x^2 + q_y^2$. This is a general
expression for all impurity potentials.

\subsection{Conductivity Formula}

The self-consistent equation of selfenergy can be written
within a compact style as
\begin{equation}
\Sigma_{0, \pm 1}^r = \frac{  \vartheta_{0, \pm 1}}{\varepsilon_F -
\Sigma_{0, \pm 1}^r},
\end{equation}
where we have denoted
\begin{equation}
\left\{
\begin{aligned}
&\vartheta_0= n_i \int [d \mathbf{q}] u_{\mathbf{q}} u_{- \mathbf{q}}
e^{- q^2 \ell_B^2 / 2},\\
&\vartheta_{\pm 1}= \frac{n_i}{4}\int [d \mathbf{q}] u_{\mathbf{q}}
u_{- \mathbf{q}} e^{- q^2 \ell_B^2 / 2}\ell_B^2 q^2.
\end{aligned}\right.
\end{equation}
The solution of this self-consistent equation gives 
\begin{equation}
\Sigma_{0, \pm 1}^r = \frac{1}{2}\left( \varepsilon_F \pm
\sqrt{\varepsilon_F^2 - 4  \vartheta_{0, \pm 1}}\right).
\end{equation}

To ensure the selfenergy has imaginary part we require $\varepsilon_F^2 <
4  \vartheta_{0, \pm 1}$. Now, the line-width function is given by
\begin{equation}
\begin{aligned}
\Delta_{0, \pm 1} &= - \text{Im} [\Sigma_{0, \pm 1}^r] = \frac{1}{2}
\sqrt{4  \vartheta_{0, \pm 1} - \varepsilon_F^2} .
\end{aligned}
\end{equation}
We only keep the positive solution of spectral function, which reads
\begin{equation}
\begin{aligned}
\mathcal{A}_{0, \pm 1} (\varepsilon_F) &= \frac{2 \Delta_{0, \pm
1}}{(\varepsilon_F - \varepsilon_{0, \pm 1})^2 + \Delta_{0, \pm 1}^2}= \frac{\sqrt{4  \vartheta_{0, \pm 1} -
\varepsilon_F^2}}{(\varepsilon_F - \varepsilon_{0, \pm 1})^2 +
(4  \vartheta_{0, \pm 1} - \varepsilon_F^2) / 4}.
\end{aligned}
\end{equation}
Consequently, from Eq.~\eqref{Eq: sigmaxx_Dirac}, the conductivity is written as
\begin{equation}
\begin{aligned}
\sigma_{xx, \pm 1}^{ql} &= \frac{e^2 v^2}{8 \pi \hbar \mathcal{V}} 
\sum_{k_x}\frac{\sqrt{4  \vartheta_0 - \varepsilon_F^2}}{\varepsilon_F^2 +
(4  \vartheta_0 - \varepsilon_F^2) / 4}\frac{\sqrt{4  \vartheta_{\pm 1} -
\varepsilon_F^2}}{(\varepsilon_F - \varepsilon_{\pm 1})^2 +
(4  \vartheta_{\pm 1} - \varepsilon_F^2) / 4}\\
&= \frac{e^2 v^2}{8 \pi \hbar}\frac{1}{2 \pi \ell_B^2}\frac{4
\sqrt{  \vartheta_0   \vartheta_{\pm 1}}}{  \vartheta_0 (\varepsilon_{\pm
1}^2 +  \vartheta_{\pm 1})}\\
&= \frac{e^2 v^2}{2 \pi h}\frac{1}{\ell_B^2}
\sqrt{\frac{  \vartheta_{\pm 1}}{  \vartheta_0}}\frac{1}{\varepsilon_{\pm
1}^2}\frac{1}{1 +  \vartheta_{\pm 1} / \varepsilon_{\pm 1}^2}\\
&= \frac{e^2 }{4 \pi h}\sqrt{\frac{  \vartheta_{\pm 1}}{  \vartheta_0}}
\frac{1}{1 +  \vartheta_{\pm 1} / \varepsilon_{\pm 1}^2},
\end{aligned}
\end{equation}
where $\varepsilon_0=0$, $\varepsilon_{\pm 1} = \pm \sqrt{2v^2/\ell_B^2}$ with $\ell_B = \sqrt{\hbar /(eB)}$,  we have set $\varepsilon_F = 0$, and the summation over $k_x$
has been interpreted as Landau degeneracy
\begin{equation}
\frac{1}{\mathcal{V}}\sum_{k_x}\rightarrow \frac{1}{2 \pi \ell_B^2} .
\end{equation}

\subsection{$\delta$-function Potential}

First, we consider the simplest case, i.e., the Fourier component of $\delta$-function potential reads
\begin{equation}
u_{\mathbf{q}} = u_0 .
\end{equation}
Thus, for $n = 0$ Landau level, we can obtain that
\begin{equation}
\begin{aligned}
\vartheta_0 & = n_i \int [d \mathbf{q}] u_{\mathbf{q}} u_{- \mathbf{q}}
e^{- q^2 \ell_B^2 / 2}\\
& = n_i u_0^2 \int \frac{q d q}{2 \pi} e^{- q^2 \ell_B^2 / 2}\\
& = \frac{n_i u_0^2}{2 \pi \ell_B^2} .
\end{aligned}
\end{equation}
Similarly, for $n=\pm 1$ Landau levels, we have
\begin{equation}
\begin{aligned}
\vartheta_{\pm 1} & = \frac{n_i}{4}\int [d \mathbf{q}] u_{\mathbf{q}}
u_{- \mathbf{q}} e^{- q^2 \ell_B^2 / 2}\ell_B^2 q^2\\
& = \frac{n_i u_0^2}{4}\int \frac{d q}{2 \pi} e^{- q^2 \ell_B^2 / 2}
\ell_B^2 q^3\\
& = \frac{n_i u_0^2}{4 \pi \ell_B^2} .
\end{aligned}
\end{equation}

We thus obtain the conductivity for the $\delta$-function potential as
\begin{equation}\label{Eq: sigma_delta}
\begin{aligned}
\sigma_{xx}^{ql, \delta} &= 2 \times \frac{e^2 }{4 \pi h}
\sqrt{\frac{  \vartheta_{\pm 1}}{  \vartheta_0}}\frac{1}{1
+  \vartheta_{\pm 1} / \varepsilon_{\pm 1}^2}\\
&= \frac{e^2 }{2 \pi h}\frac{1}{\sqrt{2}}\left( 1 + \frac{n_i u_0^2}{4
\pi \ell_B^2}\frac{\ell_B^2}{2 v^2}\right)^{- 1}\\
&= \frac{e^2}{2 \sqrt{2}\pi h}\left( 1 + \frac{n_i u_0^2}{8\pi v^2}\right)^{- 1}\\
&= \frac{e^2}{2 \sqrt{2}\pi h}\left( 1 + \gamma
\right)^{- 1},
\end{aligned}
\end{equation}
where the factor $2$ comes from the summation over $\pm$ and $\varepsilon_{\pm 1} = \pm
\sqrt{2 v^2} / \ell_B$, and we have defined a dimensionless quantity $\gamma = n_i u_{0 }^2 / (8\pi v^2)$ to characterize the strength of impurity. Considering the weak impurity limit, we approximately have
\begin{equation}
\begin{aligned}
\sigma_{xx}^{ql, \delta} &\simeq \frac{e^2}{2 \sqrt{2}\pi h} .
\end{aligned}
\end{equation}
This recovers the minimal conductivity of 2D massless Dirac fermions. 

The Hall conductivity vanishes at the charge neutral point, the longitudinal conductivity thus reads
\begin{equation}
\rho_{xx}^{ql,\delta}= \frac{1}{\sigma_{xx}^{ql,\delta}}=\frac{2 \sqrt{2}\pi h}{e^2}\left( 1 + \gamma
\right).
\end{equation}

\subsection{Gaussian Potential}

For the Gaussian potential, the Fourier component is
\begin{equation}
u_{\mathbf{q}} = u_0 e^{- \frac{q^2 \lambda^2}{2}} .
\end{equation}
Thus, for $n = 0$ Landau level, we can obtain that
\begin{equation}
\begin{aligned}
\vartheta_0 & = n_i \int [d \mathbf{q}] u_{\mathbf{q}} u_{-
\mathbf{q}} e^{- q^2 \ell_B^2 / 2}\\
& = n_i u_0^2 \int \frac{q d q}{2 \pi} e^{- q^2 \lambda^2} e^{- q^2
\ell_B^2 / 2}\\
& = \frac{n_i u_0^2}{2 \pi (\ell_B^2 + 2 \lambda^2)} .
\end{aligned}
\end{equation}
Similarly, for $n=\pm 1$ Landau levels, we have
\begin{equation}
\begin{aligned}
\vartheta_{\pm 1} & = \frac{n_i}{4}\int [d \mathbf{q}] u_{\mathbf{q}}
u_{- \mathbf{q}} e^{- q^2 \ell_B^2 / 2}\ell_B^2 q^2\\
& = \frac{n_i u_0^2}{4}\int \frac{d q}{2 \pi} e^{- q^2 \lambda^2} e^{-
q^2 \ell_B^2 / 2}\ell_B^2 q^3\\
& = \frac{n_i u_0^2\ell_B^2}{4 \pi (\ell_B^2 + 2 \lambda^2)^2} .
\end{aligned}
\end{equation}

We thus obtain the conductivity for the Gaussian potential as
\begin{equation}
\begin{aligned}
\sigma_{xx}^{ql, G} &= 2 \times \frac{e^2 }{4 \pi h}
\sqrt{\frac{  \vartheta_{\pm 1}}{  \vartheta_0}}\frac{1}{1
+  \vartheta_{\pm 1} / \varepsilon_{\pm 1}^2}\\
&=\frac{e^2}{2\sqrt{2}\pi h}\sqrt{\frac{\ell_B^2}{\ell_B^2+2\lambda^2}}\left[1+\frac{\gamma\ell_B^4}{(\ell_B^2+2\lambda^2)^2}\right]^{-1}\\
&=\frac{e^2}{2\sqrt{2}\pi h}\sqrt{\frac{1}{1+2\lambda^2/\ell_B^2}}\left[1+\frac{\gamma}{(1+2\lambda^2/\ell_B^2)^2}\right]^{-1},
\end{aligned}
\end{equation}
where the factor $2$ comes from the summation over $\pm$ and $\varepsilon_{\pm 1} = \pm
\sqrt{2 v^2} / \ell_B$, and $\gamma = n_i u_{0 }^2 / (8\pi v^2)$. When the decay length $\lambda \rightarrow
0$, we have
\begin{equation}
\sigma_{xx}^{ql, G}\simeq \frac{e^2}{2 \sqrt{2}\pi h}\left(1+\gamma
\right)^{- 1},
\end{equation}
which returns to the result of $\delta$-function potential [Eq.~\eqref{Eq: sigma_delta}].

Consequently, the longitudinal resistivity reads
\begin{equation}\label{Eq: Rho_Gauss}
\begin{aligned}
\rho_{xx}^{ql,G}&=\frac{1}{\sigma_{xx}^{ql, G}}=\frac{2\sqrt{2}\pi h}{e^2}\sqrt{1+2\lambda^2/\ell_B^2}\left[1+\frac{\gamma}{(1+2\lambda^2/\ell_B^2)^2}\right].
\end{aligned}
\end{equation}
When $\lambda<\ell_B$, the resistivity $\rho_{xx}^{ql,G}$ can be approximately expanded as
\begin{equation}
\begin{aligned}
\rho_{xx}^{ql,G}&\simeq\frac{2\sqrt{2}\pi h}{e^2}\Big(1+\frac{\lambda^2}{\ell_B^2}\Big) \left[1+\gamma\Big(1-\frac{4\lambda^2}{\ell_B^2}\Big)\right]\\
&=\frac{2\sqrt{2}\pi h}{e^2}\left[1+\gamma+(1-3\gamma)\frac{\lambda^2}{\ell_B^2}\right]\\
&=\frac{2\sqrt{2}\pi h}{e^2}\left[1+\gamma+(1-3\gamma)\frac{\lambda^2eB}{\hbar}\right],
\end{aligned}
\end{equation}
where $\ell_B = \sqrt{\hbar /(eB)}$. This result shows a linear magnetoresistivity behavior. Thus, the slope can be obtained as
\begin{equation}
\frac{d\rho_{xx}^{ql,G}}{dB}=\frac{2\sqrt{2}\pi h}{e^2}\frac{\lambda^2e}{\hbar}(1-3\gamma)=\frac{4\sqrt{2}\pi^2}{e}\lambda^2(1-3\gamma).
\end{equation}

\subsection{Yukawa Potential}

For the Yukawa potential (screened Coulomb potential), the Fourier component
is
\begin{equation}
u_{\mathbf{q}} = \frac{e^2}{2 \epsilon}\frac{1}{\sqrt{q^2 + \kappa^2}} .
\end{equation}
Thus, for $n = 0$ Landau level, we can obtain that
\begin{equation}
\begin{aligned}
\vartheta_0 & = n_i  \int [d \mathbf{q}] u_{\mathbf{q}} u_{-
\mathbf{q}} e^{- q^2 \ell_B^2 / 2}\\
& = \frac{n_i e^4}{4 \epsilon^2}\int \frac{q d q}{2 \pi}\frac{1}{q^2
+ \kappa^2} e^{- q^2 \ell_B^2 / 2}\\
& = \frac{n_i e^4}{16 \pi \epsilon^2} e^{\ell_B^2 \kappa^2/2}\Gamma_0
(\ell_B^2 \kappa^2/2),
\end{aligned}
\end{equation}
where the incomplete gamma function is defined by
\begin{equation}
\Gamma_0 (x) = \int_x^{\infty}\frac{1}{t} e^{- t} dt.
\end{equation}
Similarly, for $n=\pm 1$ Landau levels, we have
\begin{equation}
\begin{aligned}
\vartheta_{\pm 1} & = \frac{n_i}{4}\int [d \mathbf{q}] u_{\mathbf{q}}
u_{- \mathbf{q}} e^{- q^2 \ell_B^2 / 2}\ell_B^2 q^2\\
& = \frac{n_i e^4}{16 \epsilon^2}\int \frac{d q}{2 \pi}\frac{1}{q^2 +
\kappa^2} e^{- q^2 \ell_B^2 / 2}\ell_B^2 q^3\\
& = \frac{n_i e^4 }{32 \pi \epsilon^2}\Big[ 1 + \frac{1}{2} e^{\ell_B^2
\kappa^2 / 2}\ell_B^2 \kappa^2\Gamma_0(\ell_B^2 \kappa^2/2)\Big].
\end{aligned}
\end{equation}

To proceed, we need calculate the reciprocal screening radius $\kappa$ by using the RPA [Fig. 2(e) in the article]. We
start with
\begin{equation}
\begin{aligned}
\kappa (i \omega_n, q) & = - \frac{e^2}{\epsilon \mathcal{V}} 
\int_0^{\beta} d \tau e^{i \omega_n \tau}\sum_{\mathbf{k}} g_k (\tau)
g_{k + q}  (- \tau)\\
& = - \frac{e^2}{\epsilon \mathcal{V}}\sum_{\mathbf{k}}\int_0^{\beta}
d \tau e^{i \omega_n \tau}\Bigg[ \frac{1}{\beta}\sum_{ip_n} e^{- ip_n
\tau} g_{k } (ip_n) \Bigg]  \Bigg[ \frac{1}{\beta}\sum_{ip_n'}
e^{ip_n' \tau} g_{k + q} (ip_n') \Bigg]\\
& = - \frac{e^2}{\epsilon \mathcal{V}}\sum_{\mathbf{k}}\int_0^{\beta}
d \tau \frac{1}{\beta^2}\sum_{ip_n, ip_n'} e^{(i \omega_n + ip_n' -
ip_n) \tau} g_k  (ip_n) g_{k + q}  (ip_n')\\
& = - \frac{e^2}{\epsilon \mathcal{V}}\sum_{\mathbf{k}}
\frac{1}{\beta}\sum_{ip_n, ip_n'}\delta_{i \omega_n + ip_n' - ip_n}
g_k  (ip_n) g_{k + q}  (ip_n')\\
& = - \frac{e^2}{\epsilon \mathcal{V}}\sum_{\mathbf{k}}
\frac{1}{\beta}\sum_{ip_n} g_k  (ip_n + i \omega_n) g_{k + q} (ip_n)\\
& = - \frac{e^2}{\epsilon}\int [d\mathbf{k}]  \frac{1}{\beta} 
\sum_{ip_n}\frac{1}{ip_n + i \omega_n -  \xi_k}\frac{1}{ip_n -  \xi_{k +
q}}\\
& = - \frac{e^2}{\epsilon}\int [d\mathbf{k}]  \frac{n_F ( \xi_{k + q}) - n_F ( \xi_k)}{i \omega_n +  \xi_{k_z + q} -  \xi_{k_z}},
\end{aligned}
\end{equation}
where $g_k = 1 / (ip_n -  \xi_{k })$ is the free Green function with $ \xi_{k }
= \varepsilon_k - \mu$. Then, for the static screening, we simply take the limit $i \omega_n \rightarrow 0$ and $q\rightarrow 0$:
\begin{equation}
\begin{aligned}
\kappa & = \lim_{i \omega_n \rightarrow 0, q \rightarrow 0} -
\frac{e^2}{\epsilon}\int [d\mathbf{k}]  \frac{n_F ( \xi_{k  + q}) - n_F
( \xi_k)}{i \omega_n +  \xi_{k  + q} -  \xi_k}\\
& = \frac{e^2}{\epsilon}\int [d\mathbf{k}]  \left[ - \frac{\partial
n_F ( \xi_{k, 0})}{\partial  \xi_{k, 0}}\right]\\
& = \frac{e^2}{\epsilon}\int [d\mathbf{k}] \delta (\varepsilon_k -
\mu)\\
& = \frac{e^2}{2 \pi \epsilon}\int k d k \frac{1}{v}\delta (k - k_F)\\
& = \frac{e^2}{2 \pi \epsilon}\frac{k_F}{v} .
\end{aligned}
\end{equation}
In the presence of perpendicular magnetic field, $k^2$ is quantized into a
series of Landau levels as
\begin{equation}
\begin{aligned}
k_x^2 + k_{y }^2 &\rightarrow \frac{1}{\ell_B^2} (2 n + 1)
\end{aligned}
\end{equation}
with the magnetic length $\ell_B = \sqrt{\hbar/(eB)} .$ For $n = 0$ Landau
level, we thus obtain that
\begin{equation}
\begin{aligned}
\kappa &= \frac{e^2}{2 \pi \epsilon}\frac{1}{\ell_B v} .
\end{aligned}
\end{equation}

For simplicity, we define a dimensionless quantity
\begin{equation}
\begin{aligned}
\mathcal{C}=\frac{1}{2}\ell_B^2 \kappa^2= \frac{\ell_B^2}{2}\frac{e^4}{4 \pi^2
\epsilon^2}\frac{1}{\ell_B^2 v^2} = \frac{e^4}{8 \pi^2 \epsilon^2 v^2},
\end{aligned}
\end{equation}
which is independent of magnetic field $\mathbf{B}$, and we further denote
\begin{equation}
g=\Gamma_0
(\ell_B^2 \kappa^2/2)=\Gamma_0
(\mathcal{C}).
\end{equation}
Therefore, $\vartheta_0$ and $\vartheta_{\pm 1}$ become
\begin{equation}
\left\{
\begin{aligned}
&\vartheta_0=\frac{n_i e^4}{16 \pi \epsilon^2} e^{\ell_B^2 \kappa^2/2}\Gamma_0
(\ell_B^2 \kappa^2/2)=\frac{n_i e^4 }{16 \pi \epsilon^2} ge^{\mathcal{C}},\\
&\vartheta_{\pm 1}= \frac{n_i e^4 }{32 \pi \epsilon^2}\Big[ 1 + \frac{1}{2} e^{\ell_B^2
\kappa^2 / 2}\ell_B^2 \kappa^2\Gamma_0(\ell_B^2 \kappa^2/2)\Big]= \frac{n_i e^4 }{32 \pi \epsilon^2}\left(1+g e^{\mathcal{C}}\mathcal{C}\right).
\end{aligned}
\right.
\end{equation}

We now obtain the conductivity for the $\text{Yukawa}$ potential as
\begin{equation}
\begin{aligned}
\sigma_{xx}^{ql,Y} &= 2 \times \frac{e^2 }{4 \pi h}
\sqrt{\frac{  \vartheta_{\pm 1}}{  \vartheta_0}}\frac{1}{1 + \vartheta_{\pm 1} / \varepsilon_{\pm 1}^2}\\
&= \frac{e^2 }{2 \pi h}\sqrt{\frac{\frac{n_i e^4}{32 \pi \epsilon^2}
\left(1 + g e^{\mathcal{C}}\mathcal{C}  \right)}{\frac{n_i e^4}{16 \pi \epsilon^2}
ge^{\mathcal{C}}}}\left[1 +\frac{\ell_B^2}{2 v^2}\frac{n_i e^4}{32 \pi \epsilon^2}\left( 1 +
g e^{\mathcal{C}}\mathcal{C} \right) \right]^{- 1}\\
&= \frac{e^2 }{2 \sqrt{2}\pi h}\sqrt{1/(ge^{\mathcal{C}}) + \mathcal{C}}\left[ 1 + \frac{\pi n_i\ell_B^2}{8}\mathcal{C} (1 + ge^{\mathcal{C}}\mathcal{C})\right]^{-1},
\end{aligned}
\end{equation}
where the factor $2$ comes from the summation over $\pm$ and $\varepsilon_{\pm 1} = \pm
\sqrt{2 v^2} / \ell_B$.

Therefore, the longitudinal resistivity reads
\begin{equation}
\begin{aligned}
\rho_{xx}^{ql,Y}&=\frac{1}{\sigma_{xx}^{ql,Y}}=\frac{2\sqrt{2}\pi h}{e^2}\frac{1}{\sqrt{1/(ge^{\mathcal{C}}) + \mathcal{C}}}\left[ 1 + \frac{\pi n_i\ell_B^2}{8}\mathcal{C} (1 + ge^{\mathcal{C}}\mathcal{C})\right].
\end{aligned}
\end{equation}

\section{Comparison with 2D Electron Gas}
\subsection{Landau Quantization}

For a 2D electron gas (2DEG) under the constant uniform magnetic field, the Hamiltonian is 
\begin{equation}
\mathcal{H}=\frac{1}{2m}(\mathbf{p}+e\mathbf{A})^2,
\end{equation}
where $m$ is the effective mass, and $\mathbf{A}$ is the magnetic vector potential. We choose the Landau gauge
\begin{equation}
\mathbf{A}=(-yB,0).
\end{equation}
Now, the Hamiltonian reads 
\begin{equation}
\begin{aligned}
\mathcal{H}&=\frac{1}{2m}(\mathbf{p}+e\mathbf{A})^2=\frac{\hbar^2}{2m}\left[\left(k_x-\frac{eB}{\hbar}y\right)^2-\partial_y^2\right].
\end{aligned}
\end{equation}
By defining the magnetic length $\ell_B=\sqrt{\hbar/eB}$, we have
\begin{equation}
\begin{aligned}
\mathcal{H}=\frac{\hbar^2}{2m\ell_B^2}\left[\left(\frac{y-y_0}{\ell_B}\right)^2-\ell_B^2\partial_y^2\right]=\frac{1}{2}\hbar\omega\left[\left(\frac{y-y_0}{\ell_B}\right)^2-\ell_B^2\partial_y^2\right],\\
\end{aligned}
\end{equation}
where $y_0=\ell_B^2k_x$ is the guiding center, and $\omega=eB/m$ is the cyclotron frequency. We now introducing the Ladder operators as
\begin{equation}
\left\{
\begin{aligned}
a^\dagger&\equiv-\frac{1}{\sqrt{2}}\left[\frac{y-y_0}{\ell_B}-\ell_B\partial_y\right],\\
a&\equiv-\frac{1}{\sqrt{2}}\left[\frac{y-y_0}{\ell_B}+\ell_B\partial_y\right].\\
\end{aligned}
\right.
\end{equation}
It can be proved that
\begin{equation}\label{Eq: ladder}
\begin{aligned}
\left[a, a^{\dagger}\right] &=\left[-\frac{1}{\sqrt{2}}\left(\frac{y-y_{0}}{\ell_{B}}+\ell_{B} \partial_{y}\right),-\frac{1}{\sqrt{2}}\left(\frac{y-y_{0}}{\ell_{B}}-\ell_{B} \partial_{y}\right)\right] \\
&=\frac{1}{2}\left(\left[\frac{y}{\ell_{B}},-\ell_{B} \partial_{y}\right]+\left[\ell_{B} \partial_{y}, \frac{y}{\ell_{B}}\right]\right)\\
&=1.\\      
\end{aligned}
\end{equation}

Now, by rewriting the Hamiltonian within $a$ and $a^\dagger$, we arrive at
\begin{equation}
\begin{aligned}
\mathcal{H}&=\frac{1}{2}\hbar\omega\left[\left(\frac{y-y_0}{\ell_B}\right)^2-\ell_B^2\partial_y^2\right] \\
&=\frac{1}{2}\hbar\omega\left(a a^{\dagger}+a^{\dagger} a\right) \\
&=\hbar\omega\big(a^{\dagger}a+\frac{1}{2}\big),\\   
\end{aligned}
\end{equation}
which is nothing but the Hamiltonian of “harmonic oscillator”. With the help of the solutions of harmonic oscillator, we can get the eigenenergies of 2DEG under a magnetic field
\begin{equation}
\begin{aligned}
\varepsilon_n&=\hbar\omega\big(n+\frac{1}{2}\big),\quad\quad(n=0,1,2\cdots),
\end{aligned}
\end{equation}
where $\varepsilon_n$ are called Landau levels. In the $\hat{x}$ direction, the $k_x$ is still a “good” quantum number, therefore, the wavefunctions are 
\begin{equation}\label{Eq: waveF}
\psi_{k_x,n}(x,y)=\frac{e^{ik_x x}}{\sqrt{L_x}}\phi_{k_x,n}(y)=\frac{e^{ik_x x}}{\sqrt{L_x}} \frac{1}{\sqrt{n!2^n \sqrt{\pi}\ell_B}}e^{-\zeta^2/2}H_n(\zeta)\\
\end{equation}
with
\begin{equation}
\phi_{k_x,n}(y)=\frac{1}{\sqrt{n!2^n \sqrt{\pi}\ell_B}}e^{-\zeta^2/2}H_n(\zeta),
\end{equation}
where $\zeta=(y-y_0)/\ell_B$, $y_0=\ell_B^2k_x$ is the guiding center, and $H_n$ is the Hermite polynomial. The first two Hermite polynomials read
\begin{equation}
H_0=1,\quad H_1=2\zeta.
\end{equation}

For a free 2D electron gas, the dispersion relation is $\varepsilon=\hbar^2k^2/2m$. Thus, the density of states is given by
\begin{equation}
\begin{aligned}
\int \rho d\varepsilon=\int\frac{d^d \mathbf{k}}{(2\pi)^d}=\int\frac{d\varphi kdk}{(2\pi)^2}=\int\frac{kdk}{2\pi},
\end{aligned}
\end{equation}
then
\begin{equation}
\begin{aligned}
\rho d\varepsilon&=\rho\frac{\hbar^2k}{m}dk=\frac{kdk}{2\pi}.
\end{aligned}
\end{equation}
Thus, we get the density of states
\begin{equation}
\rho=\frac{m}{2\pi\hbar^2}.
\end{equation}
Now, the number of states between energy space $\varepsilon_n$ and $\varepsilon_{n+1}$ is
\begin{equation}
\begin{aligned}
N_{L} & = \int_{(n+1/2) \hbar \omega}^{(n+3/2) \hbar \omega} \rho d \varepsilon=\frac{m}{2 \pi \hbar^{2}}\left(\frac{3}{2} \hbar \omega-\frac{1}{2} \hbar \omega\right)=\frac{1}{2 \pi \ell_{B}^{2}},
\end{aligned}
\end{equation}
this is the degeneracy of the Landau levels for 2D electron gas.

\subsection{Self-consistent Born Approximation}

In the quantum limit, the Fermi level cuts the lowest Landau level. We have
assumed that the magnetic field along the $\hat{z}$ direction. Recalling the
Kubo formula for magnetotransport, the transverse conductivity is
\begin{equation}
\sigma_{xx} = \frac{e^2 \hbar}{4 \pi \mathcal{V}}\sum_{l, l'} |v_{l,
l'}^x |^2 \mathcal{A}_l (\varepsilon_F) \mathcal{A}_{l'} (\varepsilon_F),
\end{equation}
where the summation over $l, l'$ represents
\begin{equation}
\sum_{l, l'}\rightarrow \sum_{k_x, k_x'}\sum_{n, n'} .
\end{equation}
Thus, in the quantum limit ($n = 0$), the transverse conductivity can be
written as
\begin{equation}
\begin{aligned}
\sigma_{xx}^{ql} & = \frac{e^2 \hbar}{4 \pi \mathcal{V}}\sum_{k_x,
k_x'}\sum_{n'} |v_{k_x, 0 ; k_x', n'}^x |^2 \mathcal{A}_{k_x, 0}
(\varepsilon_F) \mathcal{A}_{k_x', n'} (\varepsilon_F) .
\end{aligned}
\end{equation}
To be continue, we firstly need the matrix elements for the velocity operator
in the Landau basis. The velocity operator is given by
\begin{equation}
\begin{aligned}
v_x & = \frac{1}{i \hbar}  [x, \mathcal{H} (\mathbf{k} + e \mathbf{A})]
= \frac{1}{i \hbar}\left[ x,  \frac{\hbar^2}{2m}\big( k_x - \frac{e}{\hbar} By \big)^2
+\frac{\hbar^2}{2m}k_y^2\right]=\frac{\hbar}{m}\big( k_x - \frac{e}{\hbar} By \big).
\end{aligned}
\end{equation}

By using ladder operator $a$ and $a^{\dagger}$ [Eq.~\eqref{Eq: ladder}], $v_x$ can be rewritten as
\begin{equation}
v_x=\frac{\hbar}{m}\big( k_x - \frac{e}{\hbar} By \big)=\frac{\hbar}{m}\big( k_x - \frac{1}{\ell_B^2}y \big)=\frac{\hbar}{m\ell_B\sqrt{2}}(a+a^{\dagger}).
\end{equation}
corresponding matrix elements over Landau basis read
\begin{equation}
\left\{
\begin{aligned}
& |v_{k_x, 0; k_x', 0}^x |^2 = | \langle k_x, 0| v_x |k_x', 0 \rangle
|^2 = 0,\\
& |v_{k_x, 0 ; k_x', n'}^x |^2 = | \langle k_x, 0| v_x |k_x', n'
\rangle |^2 = \frac{\hbar^2}{2 m^2\ell_B^2}n'\delta_{0,n'-1}\delta_{k_x,k_x'}.\\
\end{aligned}
\right.
\end{equation}
Thus, we have
\begin{equation}\label{Eq: sigmaxx_2DEG}
\begin{aligned}
\sigma_{xx}^{ql} & = \frac{e^2 \hbar}{4 \pi \mathcal{V}}\sum_{k_x,
k_x'}\sum_{n'} |v_{k_x, 0 ; k_x', n'}^x |^2 \mathcal{A}_{k_x, 0}
(\varepsilon_F) \mathcal{A}_{k_x', n'} (\varepsilon_F)\\
& = \frac{e^2\hbar^3}{8\pi m^2\ell_B^2\mathcal{V}}\sum_{k_x,k_x'} \sum_{n'}n'\delta_{0,n'-1}\delta_{k_x,k_x'}\mathcal{A}_{k_x, 0}
(\varepsilon_F)\mathcal{A}_{k_x, n'} (\varepsilon_F)\\
& = \frac{e^2\hbar^3}{8\pi m^2\ell_B^2\mathcal{V}}\sum_{k_x}
\mathcal{A}_{k_x,0} (\varepsilon_F)\mathcal{A}_{k_x,1}
(\varepsilon_F).
\end{aligned}
\end{equation}
The spectral functions read
\begin{equation}
\left\{
\begin{aligned}
&\mathcal{A}_{k_x, 0} (\varepsilon_F)  = \frac{2 \Delta_{k_x,
0}}{(\varepsilon_F - \varepsilon_0)^2 + \Delta_{k_x, 0}^2},\\
&\mathcal{A}_{k_x, 1} (\varepsilon_F)  = \frac{2 \Delta_{k_x, \pm
1}}{(\varepsilon_F - \varepsilon_1)^2 + \Delta_{k_x, 1}^2},
\end{aligned}\right.
\end{equation}
where the line-width function can be found as
\begin{equation}
\Delta_{k_x, n} = - \text{Im} [\Sigma_{k_x, n}^r].
\end{equation}

\subsection{Selfenergies from the Impurity Scattering}
By using self-consistent Born approximation, the selfenergy for the $n = 0$
Landau level is
\begin{equation}
\begin{aligned}
\Sigma_0^r &= \sum_{k_x', n'}\frac{| \langle k_x', n' |V|k_x, 0 \rangle
|^2}{\varepsilon_F-\varepsilon_{n'}-\Sigma_{n'}^r}.
\end{aligned}
\end{equation}
Due to the condition of quantum limit, only the $n' = 0$ term is dominant. Thus, the scattering matrix element in spatial representation is given by
\begin{equation}
\begin{aligned}
\langle k_x', 0|V|k_x, 0 \rangle & = \int d \mathbf{r}\int d
\mathbf{r}' \langle k_x', 0| \mathbf{r}\rangle \langle \mathbf{r} |V|
\mathbf{r}' \rangle \langle \mathbf{r}' |k_x, 0 \rangle\\
& = \int d \mathbf{r}\int d \mathbf{r}' \langle k_x', 0| \mathbf{r}
\rangle V (\mathbf{r}) \delta (\mathbf{r} - \mathbf{r}')  \langle
\mathbf{r}' |k_x, 0 \rangle\\
& = \int d \mathbf{r}\langle k_x', 0| \mathbf{r}\rangle V (\mathbf{r})
\langle \mathbf{r} |k_x, 0 \rangle\\
& = \int d \mathbf{r}\psi_{k_x', 0}^{\dagger} (\mathbf{r}) V
(\mathbf{r}) \psi_{k_x, 0} (\mathbf{r}),
\end{aligned}
\end{equation}
where $\langle \mathbf{r} |V| \mathbf{r}' \rangle = V (\mathbf{r}) \delta
(\mathbf{r} - \mathbf{r}')$, and the wave function of the zeroth Landau level
is [Eq.~\eqref{Eq: waveF}]
\begin{equation}
\psi_{k_x, 0} (\mathbf{r})=\frac{e^{ik_x x}}{\sqrt{L_x}}\phi_{k_x,0}=\frac{e^{ik_x x}}{\sqrt{L_x}} \frac{1}{\sqrt{ \sqrt{\pi}\ell_B}}e^{-\zeta^2/2}
\end{equation}
with $\zeta = y/\ell_B-\ell_B k_x$. Recalling the ensemble average
\begin{equation}
\langle V (\mathbf{r}) V (\mathbf{r}') \rangle_{dis} = n_i  \int [d
\mathbf{q}] e^{i \mathbf{q}\cdot (\mathbf{r} - \mathbf{r}')}
u_{\mathbf{q}} u_{- \mathbf{q}},
\end{equation}
we can obtain that
\begin{equation}
\begin{aligned}
| \langle k_x', 0| V|k_x, 0 \rangle |^2 & = \frac{n_i}{L_x^2}\int [d
\mathbf{q}] u_{\mathbf{q}} u_{- \mathbf{q}}\int d \mathbf{r}\int d
\mathbf{r}' e^{i \mathbf{q}\cdot (\mathbf{r} - \mathbf{r}')} e^{i (k_x -
k_x') x}\\
& \quad \times e^{- i (k_x - k_x') x'}\phi_{k_x, 0} (y) \phi_{k_x', 0}
(y) \phi_{k_x, 0} (y') \phi_{k_x', 0} (y')\\
& = n_i  \int [d \mathbf{q}] u_{\mathbf{q}} u_{- \mathbf{q}}\int dy
\int dy' e^{iq_y  (y - y')}\phi_{k_x, 0} (y) \phi_{k_x', 0} (y)\\
& \quad \times \phi_{k_x, 0} (y') \phi_{k_x', 0} (y') \delta_{q_x + k_x
- k_x'}\delta_{- q_x - k_x + k_x'}\\
& = n_i  \int [d \mathbf{q}] u_{\mathbf{q}} u_{- \mathbf{q}} e^{- [(k_x
- k_x')^2 + q_y^2] \ell_B^2 / 2}\delta_{q_x + k_x - k_x'},
\end{aligned}
\end{equation}
where we have used
\begin{equation}
\begin{aligned}
\int dye^{\pm iq_y y}\phi_{k_x, 0} (y) \phi_{k_x', 0} (y) & =
\frac{1}{\ell_B  \sqrt{\pi}}\int dye^{\pm iq_y y} e^{- (y / \ell_B -
\ell_B k_x)^2 / 2} e^{- (y / \ell_B - \ell_B k_x')^2 / 2}\\
& = e^{- [(k_x - k_x')^2 + q_y^2] \ell_B^2 / 4 \pm iq_y (k_x + k_x')
\ell_B^2/2}.
\end{aligned}
\end{equation}
Thus, the final result of the selfenergy for the $n = 0$ Landau level is
\begin{equation}
\begin{aligned}
\Sigma_0^r & \simeq \sum_{k_x'}\frac{| \langle k_x', 0| V|k_x, 0 \rangle
|^2}{\varepsilon_F - \varepsilon_0 - \Sigma_0^r}\\
& = \frac{n_i }{\varepsilon_F -\varepsilon_0- \Sigma_0^r}\sum_{k_x'}\int [d
\mathbf{q}] u_{\mathbf{q}} u_{- \mathbf{q}} e^{- [(k_x - k_x')^2 + q_y^2]
\ell_B^2 / 2}\delta_{q_x + k_x - k_x'}\\
& = \frac{n_i }{\varepsilon_F - \varepsilon_0-\Sigma_0^r}\int [d \mathbf{q}]
u_{\mathbf{q}} u_{- \mathbf{q}} e^{- q^2 \ell_B^2 / 2},
\end{aligned}
\end{equation}
where $q^2 = q_x^2 + q_y^2$. This is a general
expression for all impurity potentials.

The selfenergy for the $n = 1$ Landau level is given by
\begin{equation}
\begin{aligned}
\Sigma_1^r & = \sum_{k_x', n'}\frac{| \langle k_x', n' |V|k_x, \pm
1 \rangle |^2}{\varepsilon_F - \varepsilon_{n'} - \Sigma_1^r} .
\end{aligned}
\end{equation}
Due to the condition of quantum limit, only the $n' = 0$ term is dominant. The
scattering matrix element reads
\begin{equation}
\langle k_x', 0| V|k_x, 1 \rangle = \int d \mathbf{r}
\psi_{k_x', 0}^{\dagger} (\mathbf{r}) V (\mathbf{r}) \psi_{k_x, 1}
(\mathbf{r}),
\end{equation}
and the wave functions are [Eq.~\eqref{Eq: waveF}]
\begin{equation}
\left\{
\begin{aligned}
\psi_{k_x, 0} (\mathbf{r}) & = \frac{e^{ik_x x}}{\sqrt{L_x}}\phi_{k_x, 0} (y)=\frac{e^{ik_x x}}{\sqrt{L_x}}\frac{1}{\sqrt{\sqrt{\pi}\ell_B}} e^{- \zeta^2/2},\\
\psi_{k_x, 1} (\mathbf{r}) & = \frac{e^{ik_x x}}{\sqrt{2 L_x}}\phi_{k_x, 1} (y)=\frac{e^{ik_x x}}{\sqrt{2 L_x}} \frac{\sqrt{2}\zeta}{\sqrt{\sqrt{\pi}\ell_B}}
e^{- \zeta^2 / 2}
\end{aligned}
\right.
\end{equation}
with $\zeta=y/\ell_B-\ell_B k_x$. Recalling the ensemble average
\begin{equation}
\langle V (\mathbf{r}) V (\mathbf{r}') \rangle_{dis} = n_i  \int [d
\mathbf{q}] e^{i \mathbf{q}\cdot (\mathbf{r} - \mathbf{r}')}
u_{\mathbf{q}} u_{- \mathbf{q}},
\end{equation}
we can obtain that
\begin{equation}
\begin{aligned}
| \langle k_x', 0| V|k_x, 1 \rangle |^2 & = \frac{n_i}{2 L_x^2}\int
[d \mathbf{q}] u_{\mathbf{q}} u_{- \mathbf{q}}\int d \mathbf{r}\int d
\mathbf{r}' e^{i \mathbf{q}\cdot (\mathbf{r} - \mathbf{r}')} e^{i (k_x -
k_x') x}\\
& \quad \times e^{- i (k_x - k_x') x'}\phi_{k_x, 0} (y) \phi_{k_x', 1}
(y) \phi_{k_x, 0} (y') \phi_{k_x', 1} (y')\\
& = \frac{n_i}{2}\int [d \mathbf{q}] u_{\mathbf{q}} u_{- \mathbf{q}} 
\int dy \int dy' e^{iq_y  (y - y')}\phi_{k_x, 0} (y) \phi_{k_x', 1}
(y)\\
& \quad \times \phi_{k_x, 0} (y') \phi_{k_x', 1} (y') \delta_{q_x + k_x
- k_x'}\delta_{- q_x - k_x + k_x'}\\
& = \frac{n_i}{4}\int [d \mathbf{q}] u_{\mathbf{q}} u_{- \mathbf{q}}
e^{- [(k_x - k_x')^2 + q_y^2] \ell_B^2 / 2}\ell_B^2  [(k_x - k_x')^2 +
q_y^2] \delta_{q_x + k_x - k_x'},
\end{aligned}
\end{equation}
where we have used
\begin{equation}
\begin{aligned}
\int dye^{\pm iq_y y}\phi_{k_x, 0} (y) \phi_{k_x', 1} (y) & =
\frac{\sqrt{2}}{\ell_B  \sqrt{\pi}}\int dy (y / \ell_B - \ell_B k_x)
e^{\pm iq_y y} e^{- (y / \ell_B - \ell_B k_x)^2 / 2} e^{- (y / \ell_B -
\ell_B k_x')^2 / 2}\\
& = \frac{\ell_B}{\sqrt{2}} e^{- [(k_x - k_x')^2 + q_y^2] \ell_B^2 / 4
\pm iq_y  (k_x + k_x') \ell_B^2 / 2}  (- k_x + k_x' \pm iq_y) .
\end{aligned}
\end{equation}
Thus, the final result of the self-energy for the $n = 1$ Landau level is
\begin{equation}
\begin{aligned}
\Sigma_1^r & = \frac{| \langle k_x', n' |V|k_x, 1 \rangle
|^2}{\varepsilon_F - \varepsilon_0 - \Sigma_1^r}\\
& = \frac{1}{\varepsilon_F -\varepsilon_0- \Sigma_1^r}\frac{n_i}{4}
\sum_{k_x'}\int [d \mathbf{q}] u_{\mathbf{q}} u_{- \mathbf{q}} e^{-
[(k_x - k_x')^2 + q_y^2] \ell_B^2 / 2}\ell_B^2  [(k_x - k_x')^2 + q_y^2]
\delta_{q_x + k_x - k_x'}\\
& = \frac{1}{\varepsilon_F -\varepsilon_0- \Sigma_1^r}\frac{n_i}{4}\int [d
\mathbf{q}] u_{\mathbf{q}} u_{- \mathbf{q}} e^{- q^2 \ell_B^2 / 2}
\ell_B^2 q^2,
\end{aligned}
\end{equation}
where  $q^2 = q_x^2 + q_y^2$. This is a general
expression for all impurity potentials.

\subsection{Resistivity for Different Impurity Potentials}

The self-consistent equation of selfenergy can be written
within a compact style as
\begin{equation}
\begin{aligned}
\Sigma_{0, 1}^r= \frac{  \vartheta_{0, 1}}{\varepsilon_F -\varepsilon_0-
\Sigma_{0, 1}^r},
\end{aligned}
\end{equation}
where we have denoted
\begin{equation}
\left\{
\begin{aligned}
\vartheta_0 &= n_i \int [d \mathbf{q}] u_{\mathbf{q}} u_{- \mathbf{q}}
e^{- q^2 \ell_B^2 / 2},\\
\vartheta_1 &= \frac{n_i}{4}\int [d \mathbf{q}] u_{\mathbf{q}}
u_{- \mathbf{q}} e^{- q^2 \ell_B^2 / 2}\ell_B^2 q^2 .
\end{aligned}\right.
\end{equation}
The solution of this self-consistent equation gives 
\begin{equation}
\Sigma_{0, 1}^r= \frac{1}{2}\left[\varepsilon_F-\frac{1}{2}\hbar\omega \pm
\sqrt{(\varepsilon_F-\varepsilon_0)^2 - 4  \vartheta_{0, 1}}\right].
\end{equation}

To ensure the selfenergy has imaginary part we require $(\varepsilon_F-\varepsilon_0)^2 <
4  \vartheta_{0, 1}$. The line-width function now is given by
\begin{equation}
\begin{aligned}
\Delta_{0, 1} &= - \text{Im} [\Sigma_{0, 1}^r] = \frac{1}{2}
\sqrt{4  \vartheta_{0, 1} - (\varepsilon_F-\varepsilon_0)^2}.
\end{aligned}
\end{equation}
We only keep the positive solution of spectral function, which reads
\begin{equation}
\begin{aligned}
\mathcal{A}_{0, 1} (\varepsilon_F) &= \frac{2 \Delta_{0,1}}{(\varepsilon_F - \varepsilon_{0, 1})^2 + \Delta_{0, 1}^2}= \frac{\sqrt{4  \vartheta_{0, 1}-(\varepsilon_F-\varepsilon_0)^2}}{(\varepsilon_F - \varepsilon_{0, 1})^2 +\big[4\vartheta_{0, 1} - (\varepsilon_F-\varepsilon_0)^2\big] / 4}.
\end{aligned}
\end{equation}
Consequently, from Eq.~\eqref{Eq: sigmaxx_2DEG}, the conductivity is written as
\begin{equation}
\begin{aligned}
\sigma_{xx}^{ql} &= \frac{e^2\hbar^3}{8\pi m^2\ell_B^2\mathcal{V}}\sum_{k_x}
\mathcal{A}_{0} (\varepsilon_F)\mathcal{A}_{1}
(\varepsilon_F)\\
&=\frac{e^2\hbar^3}{8\pi m^2\ell_B^2} \frac{1}{2 \pi \ell_B^2}\frac{\sqrt{4  \vartheta_{0}-(\varepsilon_F-\varepsilon_0)^2}}{(\varepsilon_F - \varepsilon_{0})^2 +\big[4\vartheta_{0} - (\varepsilon_F-\varepsilon_0)^2\big]/4}\frac{\sqrt{4  \vartheta_{1}-(\varepsilon_F-\varepsilon_0)^2}}{(\varepsilon_F-\varepsilon_{1})^2 +\big[4\vartheta_{1}-(\varepsilon_F-\varepsilon_0)^2\big]/4}\\
&=\frac{e^2\hbar^3}{16\pi^2 m^2\ell_B^4}\frac{\sqrt{4 \vartheta_{0}}}{\vartheta_{0}}\frac{\sqrt{4  \vartheta_{1}}}{(\varepsilon_F-\varepsilon_{1})^2 +\vartheta_{1}}\\
&=\frac{e^2\hbar^3}{4\pi^2m^2\ell_B^4}\sqrt{\frac{\vartheta_1}{\vartheta_0}}\frac{1}{\hbar^2\omega^2+\vartheta_1}\\
&=\frac{e^2}{2\pi h}\sqrt{\frac{\vartheta_1}{\vartheta_0}}\frac{1}{1+\vartheta_1/(\hbar^2\omega^2)},
\end{aligned}
\end{equation}
where $\varepsilon_{n}=(n+1/2)\hbar\omega$ with $\omega = eB/m$, we have set $\varepsilon_F=\hbar\omega/2$, and the summation over $k_x$ has been interpreted as Landau degeneracy
\begin{equation}
\frac{1}{\mathcal{V}}\sum_{k_x}\rightarrow \frac{1}{2 \pi \ell_B^2}.
\end{equation}

Firstly, we consider the simplest case, i.e., $\delta$ potential:
\begin{equation}
u_{\mathbf{q}} = u_0 .
\end{equation}
Thus, for $n = 0$ Landau level, we can obtain that
\begin{equation}
\begin{aligned}
\vartheta_0 & = n_i \int [d \mathbf{q}] u_{\mathbf{q}} u_{- \mathbf{q}}
e^{- q^2 \ell_B^2 / 2}\\
& = n_i u_0^2 \int \frac{q d q}{2 \pi} e^{- q^2 \ell_B^2 / 2}\\
& = \frac{n_i u_0^2}{2 \pi \ell_B^2} .
\end{aligned}
\end{equation}
Similarly, for $n = 1$ Landau levels, we have
\begin{equation}
\begin{aligned}
\vartheta_1 & = \frac{n_i}{4}\int [d \mathbf{q}] u_{\mathbf{q}}
u_{- \mathbf{q}} e^{- q^2 \ell_B^2 / 2}\ell_B^2 q^2\\
& = \frac{n_i u_0^2}{4}\int \frac{d q}{2 \pi} e^{- q^2 \ell_B^2 / 2}
\ell_B^2 q^3\\
& = \frac{n_i u_0^2}{4 \pi \ell_B^2}.
\end{aligned}
\end{equation}

We thus obtain the conductivity within $\delta$ potential as
\begin{equation}
\begin{aligned}
\sigma_{xx}^{ql, \delta} &= \frac{e^2}{2\pi h}\sqrt{\frac{\vartheta_1}{\vartheta_0}}\frac{1}{1+\vartheta_1/(\hbar^2\omega^2)}\\
&=\frac{e^2}{2\sqrt{2}\pi h}\left(1+\frac{n_i u_0^2}{4 \pi \ell_B^2\hbar^2\omega^2}\right)^{-1}\\
&=\frac{e^2}{2\sqrt{2}\pi h}\left(1+\frac{n_iu_0^2m^2\ell_B^2}{4\pi\hbar^4}\right)^{-1}\\
&=\frac{e^2}{2\sqrt{2}\pi h}\left(1+\ell_B^2/\ell_e^2\right)^{-1},
\end{aligned}
\end{equation}
where we have defined a characteristic length $\ell_e=\sqrt{4\pi\hbar^4/(n_iu_0^2m^2)}$. Therefore, the longitudinal conductivity reads
\begin{equation}
\rho_{xx}^{ql,\delta}= \frac{1}{\sigma_{xx}^{ql,\delta}}=\frac{2\sqrt{2}\pi h}{e^2}\left(1+\ell_B^2/\ell_e^2\right).
\end{equation}

For the Gaussian potential, the Fourier component is
\begin{equation}
u_{\mathbf{q}} = u_0 e^{- \frac{q^2 \lambda^2}{2}} .
\end{equation}
Thus, for $n = 0$ Landau level, we can obtain that
\begin{equation}
\begin{aligned}
\vartheta_0 & = n_i \int [d \mathbf{q}] u_{\mathbf{q}} u_{-
\mathbf{q}} e^{- q^2 \ell_B^2 / 2}\\
& = n_i u_0^2 \int \frac{q d q}{2 \pi} e^{- q^2 \lambda^2} e^{- q^2
\ell_B^2 / 2}\\
& = \frac{n_i u_0^2}{2 \pi (\ell_B^2 + 2 \lambda^2)} .
\end{aligned}
\end{equation}
Similarly, for $n = 1$ Landau levels, we have
\begin{equation}
\begin{aligned}
\vartheta_1 & = \frac{n_i}{4}\int [d \mathbf{q}] u_{\mathbf{q}}
u_{- \mathbf{q}} e^{- q^2 \ell_B^2 / 2}\ell_B^2 q^2\\
& = \frac{n_i u_0^2}{4}\int \frac{d q}{2 \pi} e^{- q^2 \lambda^2} e^{-
q^2 \ell_B^2 / 2}\ell_B^2 q^3\\
& = \frac{n_i u_0^2\ell_B^2}{4 \pi (\ell_B^2 + 2 \lambda^2)^2} .
\end{aligned}
\end{equation}

We thus obtain the conductivity within Gaussian potential as
\begin{equation}
\begin{aligned}
\sigma_{xx}^{ql,G}&=\frac{e^2}{2\pi h}\sqrt{\frac{\vartheta_1}{\vartheta_0}}\frac{1}{1+\vartheta_1/(\hbar^2\omega^2)}\\
&=\frac{e^2}{2\sqrt{2}\pi h}\sqrt{\frac{\ell_B^2}{\ell_B^2+2\lambda^2}}\left[1+ \frac{n_i u_0^2\ell_B^2/(\hbar^2\omega^2)}{4 \pi (\ell_B^2 + 2 \lambda^2)^2}\right]^{-1}\\
&=\frac{e^2}{2\sqrt{2}\pi h}\sqrt{\frac{1}{1+2\lambda^2/\ell_B^2}}\left[1+\frac{n_iu_0^2m^2\ell_B^2}{4\pi\hbar^4(1+2\lambda^2/\ell_B^2)^2}\right]^{-1}\\
&=\frac{e^2}{2\sqrt{2}\pi h}\sqrt{\frac{1}{1+2\lambda^2/\ell_B^2}}\left[1+\frac{\ell_B^2/\ell_e^2}{(1+2\lambda^2/\ell_B^2)^2}\right]^{-1}.
\end{aligned}
\end{equation}
When $\ell_B \gg \lambda$ (or $\lambda \rightarrow
0$), we have
\begin{equation}
\sigma_{xx}^{ql, G}\simeq \frac{e^2}{2\sqrt{2}\pi h}\left(1+\ell_B^2/\ell_e^2\right)^{-1},
\end{equation}
which returns to the result of $\delta$ potential. Consequently, the longitudinal resistivity reads
\begin{equation}
\begin{aligned}
\rho_{xx}^{ql, G} &= \frac{1}{\sigma_{xx}^{ql, G}}=\frac{2\sqrt{2}\pi h}{e^2}\sqrt{1+2\lambda^2/\ell_B^2}\left[1+\frac{\ell_B^2/\ell_e^2}{(1+2\lambda^2/\ell_B^2)^2}\right].
\end{aligned}
\end{equation}
Notably, this result is completely different from that in Eq.~\eqref{Eq: Rho_Gauss}.

\bibliography{ref}